%% file: Main.tex
\documentclass[11pt, letterpaper]{article}

\usepackage[english]{babel}
\usepackage[utf8]{inputenc}
\usepackage[gen]{eurosym}
\usepackage{graphicx}
\usepackage{epstopdf}
\usepackage[stable]{footmisc}
\usepackage{caption}
\usepackage{subcaption} 
\usepackage[title]{appendix}
\usepackage{multirow}
\usepackage[margin=1.0in]{geometry}
\usepackage{array}
\usepackage[title]{appendix}
\usepackage{threeparttable}

\usepackage{amsmath,amssymb,amsthm,latexsym,color}
\newtheorem{definition}{\sc Definition}

\usepackage{tablefootnote}
\usepackage{lineno}
\usepackage{pifont}

\usepackage[utf8]{inputenc}
\usepackage[T1]{fontenc}
\usepackage{authblk}

\input{resources.tex}

\title{An Emissions Trading System to reach NDC targets in the Chilean electric sector}
\author[1]{P\'ia Amigo}
\author[2]{Sebasti{\'a}n Cea-Echenique}
\author[2]{Felipe Feijoo}
\affil[1]{Facultad de Ingenier{\'i}a y Ciencias, Universidad Adolfo Ib{\'a}{\~n}ez, Viña del Mar, Chile}
\affil[2]{School of Industrial Engineering, Pontificia Universidad Cat{\'o}lica de Valpara{\'i}so, Valpara{\'i}so, Chile}
\date{\today}

\begin{document}
\maketitle

\begin{abstract}
In the context of the Paris Agreement, Chile has pledged to reduce Greenhouse Gases (GHG) intensity by at least 30\% below 2007 levels by 2030, and to phase out coal as a energy source by 2040, among other strategies. In pursue of these goals, Chile has implemented a \$5 per tonne of CO$_2$ emission tax, first of its kind in Latin America. However, such a low price has proven to be insufficient. In our work, we study an alternative approach for capping and pricing carbon emissions in the Chilean electric sector; the cap and trade paradigm. We model the Chilean electric market (generators and emissions auctioneer) as a two stage capacity expansion equilibrium problem, where we allow future investment and trading of emission permits among generator agents. The model studies generation and future investments in the Chilean electric sector in two regimes of demand: deterministic and stochastic. We show that the current Chilean Greenhouse Gases (GHG) intensity pledge does not drive an important shift in the future Chilean electric matrix. To encourage a shift to greener technologies, a more stringent carbon budget must be considered, resulting in a carbon price approximately ten times higher than the present one. We also show that achieving the emissions reduction goal does not necessarily results in further reductions of carbon generation, or phasing out coal in the longer term. Finally, we demonstrate that under technology change costs reductions, higher demand scenarios will relax the need for stringent carbon budgets to achieve new renewable energy investments and hence meet the Chilean pledges. These results suggest that some aspects of the Chilean pledge require further analysis, of the economic impact, particularly with the recent announcement of achieving carbon neutrality towards 2050. 
\end{abstract}

\section{Introduction}\label{sec:intro}
Over 150 years ago, Eunice Foote, in her paper ``\textit{Circumstances affecting the Heat of the Sun's Rays}'' \cite{Foote1856} discovered for the first time in history the link between temperature and CO$_2$ and recognized the effect this link might have on the atmosphere: ``\textit{An atmosphere of that gas would give to our earth a high temperature; and if as some suppose, at one period of its history the air had mixed with it a larger proportion than as present, an increased temperature from its own action as well as from increased weight must have necessarily resulted}''. This work and other groundbreaking studies that came afterwards set the basis of climate science. To date, the unanimous scientific consensus states that Earth is warming and that the main cause are the anthropogenic Greenhouse gases (GHG) emission \cite{powell2019scientists}. One of the main ``culprits'' of polluting the environment is the energy sector, burning fossil fuels to produce heat and electricity. In fact, driven by higher energy demand in 2018, global energy-related CO$_2$ emissions reached a historic high of 33.1 GtCO$_2$ \cite{IEA2019}. Hence, the energy sector appears as a key participant in the fight towards climate crisis mitigation. This is one of the reasons why energy-sector's decarbonization strategies have been the focus of a vast number of studies, being carbon pricing the main mechanism to accomplish such a goal \cite{Robert2019Carbon}. 

\smallskip
The Paris Agreement, signed by 169 parties in 2015 in the COP21 conference, was aimed to ``keep a global temperature rise this century well below 2 $^{\circ}$C above pre-industrial levels and to pursue efforts to limit the temperature increase even further to 1.5 $^{\circ}$C''.\footnote{ https://unfccc.int/process-and-meetings/the-paris-agreement/the-paris-agreement} Participant countries submitted voluntarily their National Determined Contribution (NDC) with goals according to each economic reality\footnote{ To keep track of NDCs per country and their updated data, see https://climateactiontracker.org/ }. More than half of this NDC involved some kind of carbon pricing. In spite of the effort, it has been shown that the NDCs together are insufficient, implying a median warming of $2.6-3.1$ $^{\circ}$C by the end of the century \cite{rogelj2016paris}. This urges for stricter actions; in fact, there needs to be a massive adjustment in how the world produces its energy in order to meet climate change goals \cite{UNreport}. In this sense, experts agree that carbon pricing is one of the most economically efficient policy to reduce GHG emissions \cite{schmalensee2015lessons}.

\smallskip
There are two main types of carbon pricing, through a carbon tax and a cap-and trade system. In both cases, the main goal is to create incentives to change the investment in favor of ``green" technologies as opposed to the polluting carbon-fueled generators \cite{Chen2011}. A carbon tax puts a price on carbon, and the expected output is that emission levels will adjust accordingly; whereas a cap and trade system limits the emissions by selling allowances in a permits trading market, and the prices comes naturally from this market. Currently, the World Bank counts 46 countries which have implemented either a carbon tax or a emission trading system (ETS). In any case, there is not a clear agreement among researchers of which of the two is a more suitable climate policy option and the decision depends other aspects such as political support and administrative costs \cite{goulder2013carbon}.  One advantage for the cap-and-trade system is that it allows to set a cap (or carbon budget, as used in this paper), which can be linked to specific temperature targets \cite{feijoo2020,feijoo2019climate,Binsted2019}. 


\smallskip
Although Chile is not a major emitter of GHG, responsible only for a 0.23\% of the global emissions, the level of emissions increased dramatically a 101\% in the period 1990-2010 \cite{ChileInformeBienal2016}. The electricity generation sector is the main emitter of GHG, with a 45.1\% of the total emissions; a that is also observed globally. Motivated by this statistics and the environmental and social benefits that brings to implement mitigation policies \cite{benavides2015impact}, Chile has taken action and has made several pledge in the context of the Paris Agreement. In fact, Chile was the first country in South America to established a carbon tax of \$5 per tCO$_2$ (starting  in 2018), and, more ambitiously, it has announced carbon neutrality by 2050. Also, in their NDC, some regulations were pledged, such as phase-out coal-fueled electricity generation by 2040 and to increase electricity generation from non-conventional renewable energy (NCRE) sources up to 60\% of the total generation matrix by 2035, and 70\% by 2050. 


\smallskip

Several studies have been published about the the impact of carbon pricing and the carbon tax in Chile. There is a broad agreement that the \$5 per tCO$_2$ tax is not an efficient incentive to reduce emissions. Mardones and Flores \cite{mardones2017evaluation} performed an economic analysis to find the best emission tax, by estimating the fuel consumption and the carbon emissions. They concluded that the tax fails to generate a change in fuels, and thus, mitigate emissions, even for different costs scenarios. They estimated that a more effective tax is between \$13 - \$50 per  tCO$_2$. In addition, the economic and environmental effects of a carbon tax in Latin American countries with such implemented policy (Chile and M\'exico) and one country discussing it (Brazil) was studied by \cite{mardones2018economic}. Authors concluded that in all three countries the adopted tax is not enough to achieve the commitments made by each country in the Paris Agreement, particularly in the case of Chile. This issue is mentioned in the roadmap to 2050 document \cite{HojadeRuta}, noting that by 2030, the tax will increase to \$25 per tCO$_2$, which still is insufficient according to the studies cited above. Vera and Sauma \cite{vera2015does} also found that the Chilean carbon tax does not comply its goal, and, furthermore, they suggested that energy efficiency measures, such as introduction of flexible tariff schemes, more efficient heating system or even replacement of light and appliances, may reduce more carbon emissions than the tax, without increasing the marginal cost. 

\smallskip

More recently, authors in \cite{diaz2020equilibrium} presented an equilibrium analysis of the carbon tax in Chile. They included in their model two existing features in the chilean version of a carbon tax: pass-through restrictions and side payment rule. Both features are aimed to avoid  price raises for consumers. They find that this type of restrictions do not prevent the price increase in the long term. Moreover, concerning the ``green'' goal of the tax, investment in NCRE technologies are particularly affected, with a 79\% lower than under a ``standard'' carbon tax. 

\smallskip


Recently, the Chilean Energy Road-map established emissions guideline towards 2050, setting a target of 70\% supply from NCRE sources. In this context, authors in \cite{munoz2017aiming} noticed that the technologies that were considered as NCRE included all hydro sources, disregarding their current total capacity. This fact makes the road-map target not as restrictive as it seems. Indeed, authors showed that this target provides no incentive to renewable technologies, a result that does not change considering different fuel prices, level of demand and system configuration scenarios. Their conclusion is that the government pledge is a symbolic effort. 

\smallskip

Contrary to the tax policy, which was implemented in 2018 and have been widely studied its impact as stated above, there are fewer studies about a cap-and-trade system for Chile. The most recent example is the study from D\'iaz, Mu\~{n}oz \& Moreno \cite{diaz2020equilibrium}. They included a model with a constraint in the emission of each generator in the power market, aimed to compare a cap-and-trade system with the tax and restrictions implemented in Chile, and to asses which one is more expensive in order to achieved a desired level of carbon emissions. They find that cap-and trade system delivers a price 20\% lower than the tax regime. 

\smallskip

Mu\~{n}oz, Pumarino \& Salas \cite{munoz2017aiming} commented in their conclusions that in some cases, regulators do not performed careful analysis of the impact of policies that are implemented in their territories. In this sense, our paper is a contribution precisely in the topic of a cap-and-trade system for Chile, a system that Chile's government has manifest its interest as the Roadmap document shows, but has not performed deep analysis of its implementation or impact in their environmental policies. We model the Chilean power market including a cap-and-trade system using a two-staged capacity expansion equilibrium model \cite{ehrenmann2011stochastic}. The analysis considers a stochastic equilibrium, since we model the present operation and investment for an uncertain future, in a perfect competition market. We studied the effect of different carbon budgets (set as a cap) in the electricity production mix in Chile. The cap is estimated as the remaining carbon budget that is allowable in the electric market. We show that the emission target in Chile's NDC is not restrictive by itself, and does not change substantially the technology mix. Also, we found that the price of the carbon allowances, compared with the current tax, is greater and consistent with what previous studies have shown. In the next sections we describe the model and the assumptions made (Section \ref{sec:model}); then we give details of the Chilean electricity market and the data used (Section \ref{sec:chiledata}), and we present the results of our analysis and final remarks in Sections \ref{sec:scenarios} and \ref{sec:concl}.


\section{Model Description}\label{sec:model}

Let us consider a market of $N$ producers indexed by $i\in\{1,...,N\}$, each of them minimizing generation and investment cost. We assume perfect competition among the producers. The assumption of perfect competition is justified since in recent years several new companies have joined the Chilean electricity market, making it more competitive and without dominant firms \cite{diaz2020equilibrium,munoz2017aiming}. 

\smallskip

Our model assumes two basic stages: In the first stage (period $t=0$), electric producers decide their generation, capacity investment and carbon allowance allocation to satisfy an exogenous demand $D(0)\in\mathbb{R}_+$. In the second stage ($t\in T:=\{1,...\bar{t}\}$), uncertainty is revealed in period $t=1$. Uncertainty is represented by a state of nature $\omega\in\Omega:=\{1,..,K\}$ with probability defined by $Pr(\omega)\in[0,1]$. Thus, for a given state of nature $\omega\in\Omega$, the pair $(t,\omega)$ represents the demand in period $t>1$ when the state $\omega$ is reached. For each pair $(t,\omega)$, given an initial allowance allocation in period $t=0$ and a secondary CO$_2e$ permit trading in $t=1$, electric generators engage in a spot market and decide their generation and new capacity investment such that the exogenous stochastic demand level $D(t,\omega)\in\mathbb{R}_+^{T\times\Omega}$ is satisfied. The demand for all periods is denoted by $D=\left(D(0),(D(t,\omega))_{(t,\omega)\in T\times\Omega}\right)\in\mathbb{R}_+\times\mathbb{R}_+^{T\times\Omega}$.

\smallskip
In the first stage, $t=0$, each producer $i$ chooses the generation quantity $Q_i(t)\in\mathbb{R}_+$ such that their revenue is maximized (see Equation~\ref{eq:revenew}). The price at which each producer is paid is defined by $\pi^d(0)\in\mathbb{R}_+$. The installed capacity for each producer at $t=0$ is $\bar{Q}_i$. Each producer buys a certain amount of carbon allowances $A_i\in\mathbb{R}_+$ at a price $\pi^{a}\in\mathbb{R}_+$ from an auctioneer or regulatory agent. Finally, each producer $i$ decides the additional capacity $x_i(t)\in\mathbb{R}_+$ with a capital expenditure of $I_i\in\mathbb{R}_+$. The additional capacity becomes available after a predefined building time (in years). 

\smallskip
In the second stage, for each pair $(t,\omega)$, producer $i$ maximizes its profit by choosing the generation level $Q_i(t,\omega)\in\mathbb{R}_+$ at a price $\pi^d(t,\omega)\in\mathbb{R}_+$. We consider a technology change $TC_i(t) \in\mathbb{R}_+$ that adjusts the marginal cost of different technologies. Similarly, each producer $i$ chooses additional capacity $x_i(t,\omega)\in\mathbb{R}_+$ with an investment cost of $TCR_i(t)\cdot I_i$, where $TCR_i(t)\in\mathbb{R}_+$ is the change in the investment cost or capital expenditure over time. Additionally, we consider a permit trading system in $t=1$, where producers can purchase $P_i(\omega)\in\mathbb{R}_+$ permits from other producers if they need to surpass the initial allowances allocation $A_i$, or sell $V_i(\omega)\in\mathbb{R}_+$ unused permits. The price of the net transaction in this trading system is given by $\pi^v(\omega)\in\mathbb{R}_+$.

\smallskip
The complete nomenclature of sets, parameters and variables are summarized in Table~\ref{tab:nom}. In the following paragraphs we describe the optimization problems for each agent in this market, and we define the equilibrium concept.

\smallskip
\begin{table}
\small
    \centering
    \begin{tabular}{  l l l } 
    \hline
 \textbf{Type} &   \textbf{Units} &   \textbf{Explanation} \\ 
 \hline
 \textbf{Sets}  & &   \\ 
 $i$  & &  Producers (Technologies) \\  
 $\omega$  & & Possible scenarios \\ 
 $t$  & & Periods \\
 \hline
 \textbf{Parameters}  &  &  \\
 $I_i$  & USD/MW & Expansion cost per technology $i$ \\  
 $\bar{Q}_i$ & MW & Present operation capacity  per techonlogy $i$ \\ 
 $C_i$ &  USD/MWh & Operation cost per technology $i$ \\
 $D(0)$ &  MWh & Total demand in first stage \\
 $D(t,\omega)$ &  MWh & Total demand in period $t > 0$ (second stage) \\
 $\varepsilon_i$ &  tCO$_2e$/MWh  & Emission factor of technology $i$\\
 Pr$_\omega$ &  & Physical probability of scenario $\omega$ \\
 $\mu$ &  tCO$_2e$ & Mean of the normal distribution of the carbon budget \\
  $\sigma$ &  tCO$_2e$ & Standard deviation of the normal distribution of the carbon budget \\
  $\epsilon$ &  & Margin for total emission allowances \\
  $R$ & & Discount factor \\
  $TC_i(t)$ &  & Change in the cost of operation per technology \\
  $TCR_i(t)$ & & Change in the investment cost per technology \\
  $CF_i$ &  & Capacity factor per technology\\
  $\tau$ &  hours&  Number of hours in a year, $\tau=8760$ hours\\
   $RP_i$  &   MW &  Resource potential per technology \\
   $\textrm{lag}_i$ & years & Number of years that takes a plant to be fully operational per technology\\
\hline
 \textbf{Variables}   &  &  \\ 
 $Q_i(0)$ & MWh & produced quantity in period $t = 0 $ for producer $i$\\
 $Q_i(t,\omega)$ & MWh & produced quantity in period $t > 0 $ for producer $i$ in scenario $\omega$ \\  
 $A_i$ & tCO$_2e$ & Emission allowances purchased by producer $i$ \\ 
 $P_i(\omega)$ & tCO$_2e$ & Purchased permits in trading market by producer i  \\
 $V_i(\omega)$ & tCO$_2e$ & Sold permits in trading markets \\
 $\pi^{a}$ &  USD/tCO$_2e$  & Price of the allowances offered by the auctioneer \\
 $\pi^{v}(\omega)$ &  USD/tCO$_2e$  & Price of the permits in trading market \\
  $\pi^d(0)$ &  USD/MWh  & Price of electricity \\
 $\pi^d(t,\omega)$  & USD/MWh  & Price of electricity \\
 $x_i(0)$ &  MW  & Capacity expansion decision for period $t=0$ of producer $i$\\
 $x_i(t,\omega)$ &  MW  & Capacity expansion decision in period $t$ of producer $i$ in scenario $\omega$\\
  $\theta$ &  tCO$_2e$ & Emission allowances available in the market by the auctioneer\\
\hline
\end{tabular}
    \caption{Description of the complete nomenclature used in our model.}
    \label{tab:nom}
\end{table}

\vspace{0.5cm}

\normalsize

\subsection{Producer's problem}
Each producer $i$ represents one and only one technology in the economy (see Section~\ref{sec:chiledata}). The choice set of the producer $i$ over a time horizon of $\bar{t}$ years is given by: a) an amount of capacity expansion $x_i:=\left(x_i(0),(x_i(t,\omega))_{(t,\omega)\in T\times\Omega}\right)\in\mathbb{R}_+\times\mathbb{R}_+^{T\times\Omega}$, b) a production plan $Q_i:=\left(Q_{i}(0),(Q_{i}(t,\omega)_{(t,\omega)\in T\times\Omega})\right)\in\mathbb{R}_+\times\mathbb{R}_+^{T\times\Omega}$, c) allowances $A_i\in\mathbb{R}_+$ bought in the first period $t=0$, d) allowances $P_i(\omega)\in\mathbb{R}_+^{\Omega}$ bought in $t=1$ for time interval $t\in[1,\bar{t}]$  and e) allowances $V_i(\omega)\in\mathbb{R}_+^{\Omega}$ sold in $t=1$ for time interval $t\in[1,\bar{t}]$.

\smallskip
The model runs in one year intervals, hence, we represent the year in blocks of $\tau=8760$ hours. For each technology we consider a capacity factor $CF_i\in\mathbb{R}_+$ that represents the real operation of each plant.

The objective of producers is to minimize cost. Thus for a representative producer, there is a cost of production for each period $t\in T$. We define the revenue function given electricity prices $\pi^d:=\left(\pi^d(0),\left(\pi^d(t,\omega)\right)_{(t,\omega)\in T\times\Omega}\right)\in\mathbb{R}_+\times\mathbb{R}_+^{T\times\Omega}$ (denoted $p$ in the general form described in equation \ref{eq:revenew}) and parameters $(a_i,b_i)_{i\in\{1,...N\}}\in(\mathbb{R}^2_+)^N$ by

\begin{align}\label{eq:revenew}
f_i(p,q)=\Big(a_i\cdot q+\frac{b_i}{2}\cdot q^{2}\Big)-p\cdot q. 
\end{align}

Defining $T_0:=\{0\}\cup T$, the optimization problem of producer $i$ is given by choosing \linebreak$(x_i,Q_i, A_i,P_i,V_i)\in\mathbb{X}:=\left(\mathbb{R}_+\times\mathbb{R}_+^{T\times\Omega}\right) \times\left(\mathbb{R}_+\times\mathbb{R}_+^{T\times\Omega}\right) \times \mathbb{R}_+\times\mathbb{R}_+^{\Omega}\times\mathbb{R}_+^{\Omega}$ given prices \linebreak$(\pi^d$, $\pi^a$ , $\pi^v)\in\Pi:=\left(\mathbb{R}_+\times\mathbb{R}_+^{T\times\Omega}\right)\times\mathbb{R}_+\times\mathbb{R}^{\Omega}_+$ and parameters \linebreak$\left((a_i,b_i),I_i, TC_i(t,\omega), TCR_i(t,\omega), CF_i,\bar{Q}_i, RP_i , \varepsilon_i\right)\in \Xi:=\mathbb{R}_+^2\times\mathbb{R}_+^7$, $\tau\in \mathbb{R}_+$ and probability $(Pr(\omega))_{\omega\in\Omega}\in\Delta:=\left\{\left(Pr(\omega)\right)_{\omega\in\Omega}\in[0,1]^K:\sum_{\omega\in\Omega}Pr(\omega)=1\right\}$ as a solution of

\small
\begin{align}
\min_{(x_i,Q_i,A_i,P_i,V_i)\in \mathbb{X}}  & f_i \big( \pi^d(0),Q_i(0)\big)+ A_i \pi^{a} + I_i x_i(0) \nonumber \\ 
& + \sum_{\omega} Pr(\omega)   \Bigg[ \sum_{t>0} \frac{1}{(1+R)^t} \Big[ TC_i(t)\cdot f_i \big( \pi^d(t,\omega),Q_i(t,\omega) \big) \nonumber\\
&  + TCR_i(t) \cdot I_i\cdot x_i(t,\omega) \Big] + \pi^v(\omega)\cdot \big(P_i(\omega)-V_i(\omega)\big) \Bigg]   \label{eq:prod}
\end{align}
\begin{align}
&\textrm{subject to \ } \nonumber\\
    &\Big(CF_i \cdot\tau\Big)  \Bigg[\bar{Q}_i + x_i(0)+\sum_{t^{\prime}<t-lag_i} x_i(t^\prime,\omega) \Bigg] - Q_i(t,\omega) & \geq 0  & \qquad \forall \quad i,\omega, t  > 0   & \quad (\alpha_{i,\omega,t})      \label{eq:c1}\\
    &\Big(CF_i\cdot\tau \Big)\bar{Q_i}-Q_{i}(0)                                                                                 & \geq 0  & \qquad \forall \quad i                  &  \quad (\kappa_i) \label{eq:c2} \\
    &RP_i - \bar{Q}_i - x_i(0) - \sum_{t > 0} x_i(t,\omega)                                                                     & \geq 0  & \qquad \forall \quad i,\omega           &   \quad (\psi_{i,\omega})\label{eq:c3} \\
    &A_{i} -V_i(\omega)                                                                                                         & \geq  0 & \qquad \forall \quad i,\omega           & \quad (\beta_{i,\omega}) \label{eq:c4}\\
    &A_{i} + (P_i(\omega) - V_i(\omega))-\sum_{t>0}Q_i(t, \omega)\cdot \varepsilon_{i}-Q_i(0)\varepsilon_{i}           & \geq  0 & \qquad \forall \quad i, \omega          & \quad (\gamma_{i,\omega}) \label{eq:c5}
  \end{align}
\normalsize

Constraint~\ref{eq:c1} represents the generation capacity constraint for each producer $i$ in scenario $\omega$ and time $t$ in the second stage. Similarly, Constraint~\ref{eq:c2} restricts generation for $t=0$ (first stage). Additionally, we limit the total capacity (installed and new investments) for producer $i$ to the resource potential for each technology, as shown in Constraint~\ref{eq:c3}. The maximum amount of allowances that each generator can sell is limited by CO$_2e$ permits bought from the auctioneer (Constraint~\ref{eq:c4}) and the total emissions of each producer is restricted to all the transactions of allowances in the market (Constraint~\ref{eq:c5}). Variables in parenthesis next to each constraint represent the respective Lagrange multipliers.

\subsection{Auctioneer's problem:}\label{sec:auctioneer}
 The auctioneer chooses the total number allowances available in the market $\theta\in \mathbb{R}_+$ and sells them at a price $\pi^a$ in stage $t=0$. The price is set such that it clears the CO$_2e$ permit market among generators and the auctioneer.  The carbon budget (maximum level of CO$_2e$ emissions allowed in the second stage $t\in T$) is denoted by  $CAP\in\mathbb{R}_+$ and is drawn from a normal distribution, i.e., $CAP \thicksim N(\mu, \sigma^2)$. Hence, the variable $\theta$ must not probabilistically surpass the CO$_2e$ budget for emissions, as shown in Equation~\ref{eq:probcap}.
 
\begin{equation}
    Pr(\theta \geq CAP) \leq \epsilon \label{eq:probcap}
\end{equation}

 The objective of the auctioneer is to maximize its benefit. Note that benefit is defined as the revenue obtained by selling allowances while considering negative outcomes modeled by $\mathcal{F}(\theta)$. For instance, $\mathcal{F}(\theta)$ can represent the cost of an additional tonne of CO$_2$ imposed to the society, known as the Social Cost of Carbon (SCC)~\cite{feijoo2014design}.\footnote{The Chilean Ministry of Social Development estimated the cost of carbon for Chile to be equivalent to USD \$32 \cite{ChileSCC}.} Thus, the optimization problem for the auctioneer is given by choosing $\theta$ given price $\pi^a$ and parameters $\sigma$, $\mu$ and $\epsilon$ as a solution of:

\begin{align}
    \min_{\theta} & -\theta \pi^{a}  + \mathcal{F}(\theta)   \label{eq:auc} \\
    \textrm{s.t \ } &  \phi^{-1}(\epsilon) \sigma + \mu - \theta  \geq 0 \label{eq:au_c1}
\end{align}

where Equation~\ref{eq:au_c1} is the constraint that appears from solving for $\theta$ in Equation~\ref{eq:probcap}, where $\phi^{-1}$ is the inverse of the cumulative function of the standard normal distribution. 

\subsection{Equilibrium and solution technique}\label{sec:equi}

The data that parameterize the capacity investment model is given by the tuple $$\left(\tau,(Pr(\omega))_{\omega\in\Omega},\left(CAP,\mu,\sigma,\epsilon\right),\left((a_i,b_i),I_i, TC_i(t,\omega), TCR_i(t,\omega), CF_i,\bar{Q}_i, RP_i , \varepsilon_i\right)_{i\in N}\right)\in \mathbb{R}_+\times \Delta\times\mathbb{R}_+^4\times\Xi^N,$$

where the equilibrium notion is given by the following definition.

\begin{definition}
An equilibrium in the capacity investment model is a vector of prices and production decisions $$\left((\pi^{d*},\pi^{a*},\pi^{v*}),(x_i^*,Q_i^*,A_i^*,P_i^*,V_i^*)_{i\in\{1,...,N\}},\theta^*\right)\in\Pi\times\mathbb{X}^N \times \mathbb{R}_+$$ such that:

\begin{enumerate}
    \item $(x_i^*,Q_i^*,A_i^*,P_i^*,V_i^*)$ minimizes the cost for each producer  $i\in\{1,...,N\}$, solving problem~(\ref{eq:prod}), subject to Equations~(\ref{eq:c1})-(\ref{eq:c5}),
    \item $\theta^*$ minimizes the auctioneer cost, solving the problem in Equations~(\ref{eq:auc})-(\ref{eq:au_c1}),
    \item Market clearing conditions are satisfied:
    
\begin{align}
\textrm{(available allowances $t=0$)}: &  \ \   \sum_{i} A_{i}^* = \theta   &  & \ \  (\pi^{a*})\\
\textrm{(equilibrium in trading market $t>0$)}: &   \ \  \sum_{i} P_{i,\omega}^* = \sum_{i} V_{i,\omega}^* & \forall \ \omega & \ \ \left(\pi^{v*}(\omega)\right) \\
\textrm{(fulfillment of the demand --first stage)}:  &   \ \  \sum_{i} Q_i(0)^* = D(0), &  & \ \ (\pi^{d*}(0))\\
\textrm{(fulfillment of the demand --second stage)}:  &   \ \  \sum_{i} Q_i(t,\omega)^* = D(t,\omega), & \forall \ \omega, t& \ \ (\pi^{d*}(t,\omega))
\end{align}
\end{enumerate}
\end{definition}

The proposed equilibrium satisfies the complementarity conditions presented in Appendix~\ref{ap:mcp}, which poses the problem as a Mixed-Complementarity Problem (MCP) \cite{murphy2016tutorial,feijoo2016north}. The resulting MCP is implemented in GAMS and solved using the PATH solver \cite{Ferris2000}.

\section{Model calibration to the Chilean Electric Sector and carbon budget estimates}

In this section we present operation of the Chilean electric sector. Section~\ref{sec:chiledata} gives a description of the operation of the Chilean market and explains the method used to calibrate our model with real data. Next Section~\ref{sec:cap} describes the determination of the carbon budget based on the Chilean pledges. Finally, Section~\ref{sec:basecase} presents the calibration of the model using up-to-date data, and compares it with the real operation.

\begin{figure}[ht!]
 \includegraphics[width=\textwidth]{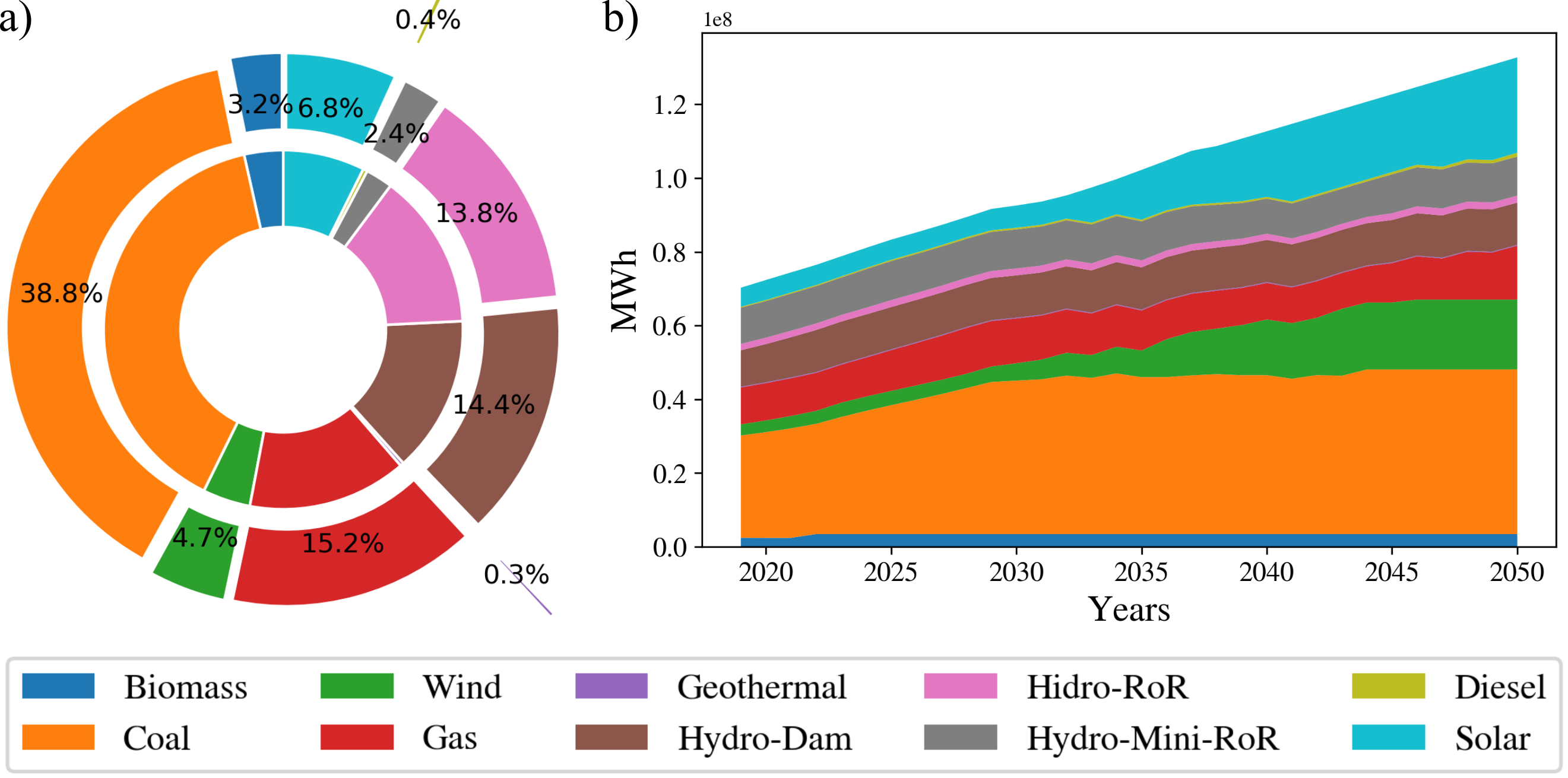}
 \caption{Panel (a): The outer ring shows the production percentage over the total electricity production in 2018 for each technology  obtained from the calibration of our model. The inner ring shows the real operation data for that year. Panel (b): Production levels per technology in the period 2019-2050 for our base-case scenario in our model, that is, business-as-usual, no green policies planned or implemented. }
  \label{fig:bc}
  \end{figure}

\subsection{Chilean electric market mix estimates}\label{sec:chiledata}

The \textit{Sistema El\'ectrico Nacional} (SEN) is the main (sub) system of the Chilean electric market\footnote{In fact, there are 2 other systems, SEA (\textit{Sistema El\'ectrico de Ays\'en}) and SEM (\textit{Sistema El\'ectrico de Magallanes}), but together they account  for less than 1\% of the total installed capacity, thus we neglected them for the purpose of this study.},  with a total generation capacity of 24.17 GW\footnote{http://energiaabierta.cl/visualizaciones/capacidad-instalada/}. For this study, we considered the existing generation units in SEN according to their main generation technology\footnote{Note that the model structure is general enough that it allows to model the electric market at a firm generation level.}: Biomass, Coal, Wind, Gas, Geothermal, Hydro-Dam, Hydro-Run-of-River (RoR), Mini Hydro-RoR (capacity less than 20 MW), Diesel and Solar (Photovoltaic)\footnote{ Although it has been announced the installation of the first concentrated solar power (CSP) plant in Chile, by the date of this study the operation has not begun and thus we neglect it. }.

\smallskip

For each generation technology, we obtained the data for capital investment and operation cost (CAPEX and OPEX), installed capacity and future planned capacities expansion through the \textit{Comisi\'on Nacional de Energ\'ia} (CNE) website. Information about the resource potentials were obtained from \cite{Santana2014potencial}.
In the case of the demand, we also use the projection provided by the CNE\footnote{http://datos.energiaabierta.cl/dataviews/251663/proyeccion-de-demanda-en-gwh/}, which foresees the demand until year 2038. This data is linearly extrapolated to extend the demand projection up to 2050. This demand  is the one used in the deterministic analysis.

\smallskip

The model is calibrated to the above data, considering the operation in 2018. This calibration was performed by fine-tuning a subset of the parameters of the model, such as cost and the technology/investment change, $TC$ and $TCR$ (see Section~\ref{sec:model} for the details). In any case, the estimation of these parameters follows the trends expected for future years (as presented by \cite{Mena2019csp,NRELcosts}), where investment and operation costs of NCRE sources are expected to decrease, as opposed to fuel-based generation where costs tends to remain constant \cite{Mena2019csp,NRELcosts}.
Figure~\ref{fig:bc}(a) shows a contrast of the electric generation mix produced in our model for 2018 with respect to the real operation data. Most of the electricity production of the country comes from coal and gas-fueled generators. Only 18\% of the energy matrix comes from NCRE sources, and taken into account large hydropower plants, this percentage increases up to  46\%. This values are in good agreement with the real operation data from CNE \cite{CNE2018}, which estimates that the contribution from NCREs is 20\% and raises to 50\% when large hydropower plants are included. 

Figure ~\ref{fig:bc}(b) shows the calibrated production level per technology over the studied period. Generation from NCRE, particularly solar and wind, does increase in future years, but still the fraction of solar and wind combined is less than 40\% by 2050, hence, coal and gas-fueled generators still prevail up to 2050. Electric generation from solar and wind technologies increase over time mainly due to the price structure (CAPEX and OPEX) which make these technologies more competitive. This business as usual (BAU) scenario is in good agreement with Chilean official projections of a BAU scenario \cite{HojadeRuta}. The electric production mix by 2050 observed in \cite{HojadeRuta} and our model are compared in Table~\ref{tab:bc_comp}. There is an small discrepancy regarding the hydro sources, with our model producing slightly less than \cite{HojadeRuta}. The reason is that hydroelectric plants depend on environmental factors such as droughts and also political. Hence, and since that in the last years some hydro plant projects have been controversial, we assume to favor NCRE sources over new installment of Hydro sources.


\begin{table}[!htbp]
\centering
  \begin{threeparttable}[t]
       \begin{tabular}{ c c c c } 
\hline
 & Technology & Base Case Roadmap 2050 & This work \\
\hline
\multirow{4}{8em}{Energy Production (\%)} & NCRE\tnote{a} & 22\%-67\% & 36.5\% \\ 
 & Hydro generation & 24\%-32\% & 18\% \\ 
& Thermal generation & 9\%-46\% & 45.5\% \\
\hline
\end{tabular}
     \begin{tablenotes}
     \footnotesize{
     \item[a] According to Chile's Energy Policies document, NCRE are Solar Photovoltaic and Concentrated Solar Power (CSP), Geothermal and Wind .
     }
   \end{tablenotes}
    \end{threeparttable}%
    \caption{Energy Generation from our model compared with Chile's Energy Policies projections for the year 2050}
  \label{tab:bc_comp}%
\end{table}%

\subsection{Determination of the carbon budget}\label{sec:cap}

The Nationally Determined Contribution (NDC) proposed by Chile states that the total GHG intensity by 2030 must be 30\% below 2007 GHG intensity of GDP, or, in absolute terms of CO$_2e$, by 2030 the emission level must be 131 MtCO$_2e$. Based on this claim, we computed the remaining emission budget (or cap) for our study period (2019 - 2050). 
\smallskip

\begin{figure}[ht!]
\centering
 \includegraphics[width=5in]{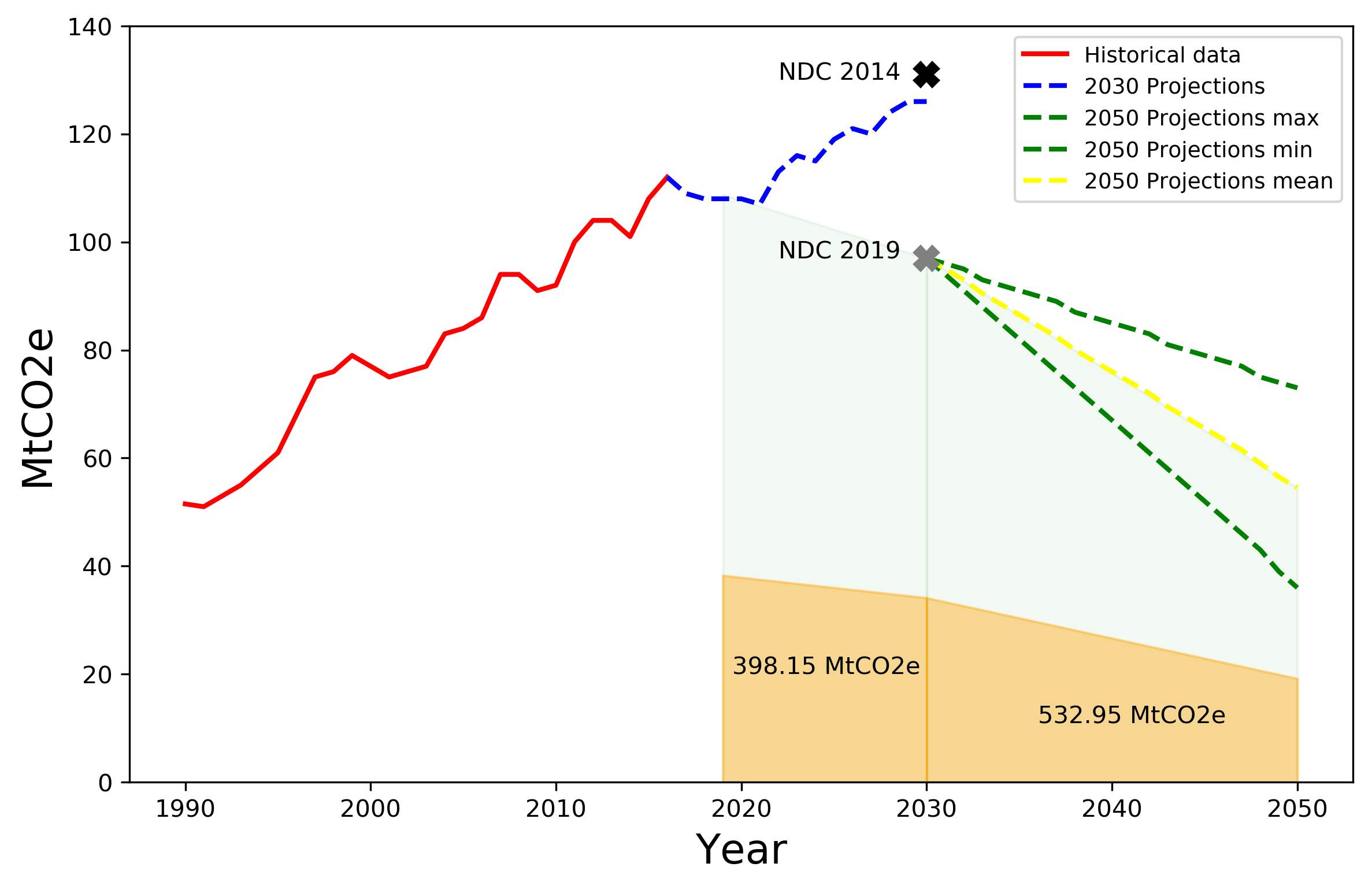}
 \caption{Emission data for Chile from Climate Action Tracker website, historical and projected. See text for details}
  \label{fig:cap}
  \end{figure}

Based on Chile's pledge in the Paris Agreement, the projected total emissions, excluding land use, land-use change and  forestry (LULUCF), up until the year 2030 are as shown in Figure~\ref{fig:cap}, based on the most recent data about Chile's NDCs found in the Climate Action Tracker (CAT) (2 Dec 2019) webpage.\footnote{https://climateactiontracker.org/countries/chile/} It should be noted that these emissions consider the main GHG, i.e, $CO_2$, $CH_4$ and $N_2O$.  In the Chilean case, the total emissions are dominated by $CO_2$ (78\%), followed by $CH_4$ with 12.5\% and $N_2O$  with 6\% \cite{ChileInformeBienal2016}. The remaining emissions are due to fluoride gases, which we will neglect in the following analysis. To account for all the GHGs, we used emission factors $\epsilon_i$ in units of carbon equivalent tCO$_2e$ taken from \cite{IPCC2014}.
\smallskip

According to \cite{ChileInformeBienal2016}, the electric generation sector was responsible for approximately 35.1\% (36.25 MtCO$_2e$) of the total emissions in 2016 (45.3\% of the emissions from the energy sector). We assume that this fraction of emissions of the electricity sector remains constant throughout the entire study period (2019-2050). This assumption is a simplification because it is expected that new green technologies will appear in the future and the fraction of CO$_2e$ emissions will be reduced. Therefore, our carbon budget determination should be taken as a upper bound. To examine the effect of this assumption, Section~\ref{sec:scenarios} shows the results of a sensitivity analysis in terms of the carbon budget (CAP), varying from 100 to 1000 MtCO$_2e$, hence, the change in the fraction of CO$_2e$ emissions from electricity sector is somewhat included when a more restrictive carbon budget is considered. 

To estimate the electricity sector carbon budget, we can refer to Figure~\ref{fig:cap}. The shaded green area in Figure~\ref{fig:cap} shows the economy wide (excluding land use emissions) carbon budget. Considering the 2019 fraction of emissions from the electric to be the same as that of 2016 (30\% of total emissions, based on \cite{ChileInformeBienal2016}), and decreasing linearly from this point towards the 2030 pledge, the shaded orange area in Figure~\ref{fig:cap} represents the electricity sector carbon budget associated with the Chilean pledge for the period 2019-2030 (398.15 $MtCO_2e$). Furthermore, recently in 2019, Chile announced its goal of becoming carbon neutral by 2050. According to the projections made by CAT, this would translate to an economy wide emissions level (by 2050) in the range of 37 to 66 $MtCO_2e$. Using the same method as before (assuming that 30\% of the emissions are associated to the electric sector), we interpolate between the 2030 emissions level and the mean estimate for 2050 (yellow line in Figure~\ref{fig:cap}). This result in an additional carbon budget of 532.95 $MtCO_2e$ (orange shaded area). Hence, the electricity sector carbon budget for the 2019-2050 period is estimated to approximately be 930 $MtCO_2e$ (398.15 $MtCO_2e$ for the 2019-2030 period plus 532.95 $MtCO_2e$ for the 2030-2050 period). We will refer to this estimate as the \textit{pledge cap}. Once again, we note that this value is an upper limit, since we made the assumption that the fraction of emission contribution from electric sector remains unchanged over time.


\subsection{Chilean pledges and scenario analysis}
\label{sec:basecase}
We explore possible scenarios that can lead to a change in the environmental policies in the country. We perform a sensitivity analysis of the carbon budget and compare the results with the planned policies and pledges that Chile has already committed in the Paris Agreement and other instances. Such analysis is needed to better estimate the necessary budget and carbon prices that Chile requires, as current implemented policies are not sufficient. These pledges are:

\begin{itemize}
    \item By 2024: retirement of the eight oldest coal-fired power plants (20\% of current coal electricity capacity). 
    \item By 2030: 30\% below 2007 GHG intensity of GDP (uncondicional target). This translates in 151\% above 1990 emissions excl. LULUCF by 2030. 
    \item By 2035: 60\% electricity production from renewable energy 
    \item By 2040: phase-out coal 
    \item By 2050: 70\% electricity production from renewable energy
\end{itemize}

\bigskip

\section{Results and discussion}\label{results}
We first study the role of the carbon budget on the evolution of the energy mix in the Chilean electric sector. As described earlier, the carbon budget is included as an emissions trading system (cap-and-trade) where generators must decide the initial number of allowances to buy and then, in future periods, generators are allowed to trade such initial allowance allocation. The cap is based on the carbon budget estimate described in Section \ref{sec:cap}. In this initial analysis, no demand uncertainty is considered, hence generators know what the demand level will be towards 2050. We then consider a stochastic analysis with five distinct (and equally likely) future demand scenarios \cite{HojadeRuta}. Hence, generators must decide on investment decisions and allowance trading strategies considering the expectation of the demand. 

\subsection{Role of the cap-and-trade system in a deterministic context}\label{sec:deterministic}

We now present the effect of an emission cap and the associated emissions trading system in the context of the Chilean electric sector under a deterministic future demand. We perform a sensitivity analysis on the carbon budget estimate presented in Section  \ref{sec:cap}. Note that the mathematical formulation of the Auctioneer's problem allows in its general form to have Normal distributed carbon budget (see $CAP$ in Section \ref{sec:auctioneer}), with mean $\mu$ and variance $\sigma^2$. For simplicity of the analysis, we do not consider a variance ($\sigma=0$) and hence we simply vary the mean $\mu$ parameter between  100 MtCO$_2e$ and 1000 MtCO$_2$e. The pledge of carbon budget determined in Section~\ref{sec:cap} (930 MtCO$_2e$) belong to that interval.

\smallskip

\begin{figure}[hbtp]
\includegraphics[height=\textheight]{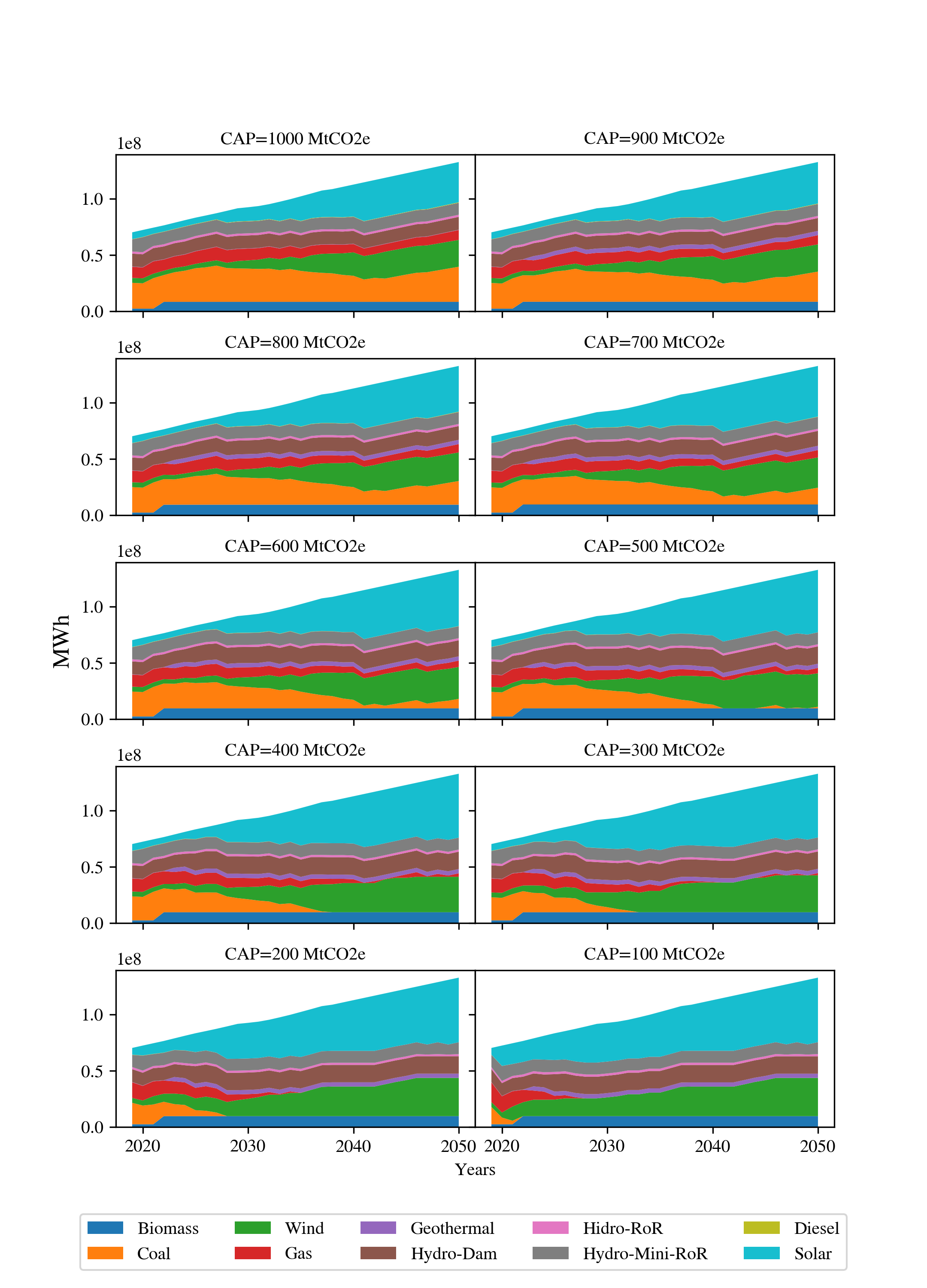}
 \caption{Production levels per technology for the period 2019-2050 at different emission caps}
  \label{fig:prod_cap}
\end{figure}
  
Figure~\ref{fig:prod_cap} shows production levels per technology towards 2050, for the carbon budget ranging between 100 MtCO$_2e$ and 1000 MtCO$_2e$. The two top panels in  this figure are the closest values to the pledge cap. Generation from conventional fossil-fuel (coal and gas) does not yield in favor of NCREs; moreover, fossil fuels are still important players in the power generation. This result shows that the Chilean pledge of lowering emissions is not sufficient by itself. According to our model, there is little economic incentive (or cost, measured by the price of carbon) for producers to invest in ``greener" technologies with a non-restrictive carbon budget. Indeed, the $CO_2e$ price obtained in the less stringent case (1000 MtCO$_2e$) is $\$24/tCO_2e$ versus a price of $\$174/tCO_2e$ in the most stringent case (100 MtCO$_2e$), representing an increment of 6.25 times in the price of carbon permits.\footnote{Furthermore, the carbon permit price of $\$24 tCO_2e$ is above the current carbon tax of $\$5 tCO_2$ implemented by the Chilean government.} 
\smallskip

The net capacity addition in the base case over the period 2019-2050 accounts to 17.38 GW, of which solar and wind account for most of it (98\%). In the case of a lax carbon budget (1000 MtCO$_2e$), total capacity additions increase to 25.58 GW, an increment of 47\% compared to the base case. New solar capacity accounts to 14.41 GW (48\% increase) whereas wind energy investment was of 9.91 GW (32\% increase). However, when an extreme (low) carbon budget is considered (100 MtCO$_2e$), solar and wind investments account to 24.67 GW and 14.94 GW (154\% and 99\% increase), respectively. We also observed that new investment does not vary significantly for carbon budgets values between 100 MtCO$_2e$ and 500 MtCO$_2e$. In fact, total new investments increase by 7\% for a budget of 100 MtCO$_2e$ when compared to 500 MtCO$_2e$. However, the allowance price in these two cases is significantly different. For a carbon budget of 500 MtCO$_2e$ the allowance price is $\$45/tCO_2e$, almost four times lower than the allowance price for 100 MtCO$_2e$ ($\$174/tCO_2e$).
\smallskip

Results also show that allowance prices lower than $\$45/tCO_2e$ are not sufficient to phase out coal towards 2040 (nor 2050), as one of the objectives that Chile has pledged. Also, it is not clear if the Chilean government is working towards this goal. In 2019, the biggest coal power plant (Mejillones) with a total capacity of 375 MW started operation. However, to date, only 2 plants have been closed, both of them accounting to 170 MW of capacity. In this sense, as the CAT website notes in their country's profile, Chile has given mixed signals about coal-fueled plants and their plan to remove them from the production matrix. It is clear from Figure~\ref{fig:prod_cap} that coal can be phased out when a carbon budget between 300-400 MtCO$_2e$  is considered, with allowance prices of $\$56/tCO_2e$ and above (as the carbon budget becomes more stringent). Still, a carbon budget of 700 MtCO$_2e$ is more restrictive than the current pledge, without imposing complete shutdown of some generation plants, which has economic implications, particularly for developing countries \cite{Binsted2019}.

\smallskip

\begin{figure}[ht!]
\centering
 \includegraphics[width=4.5in]{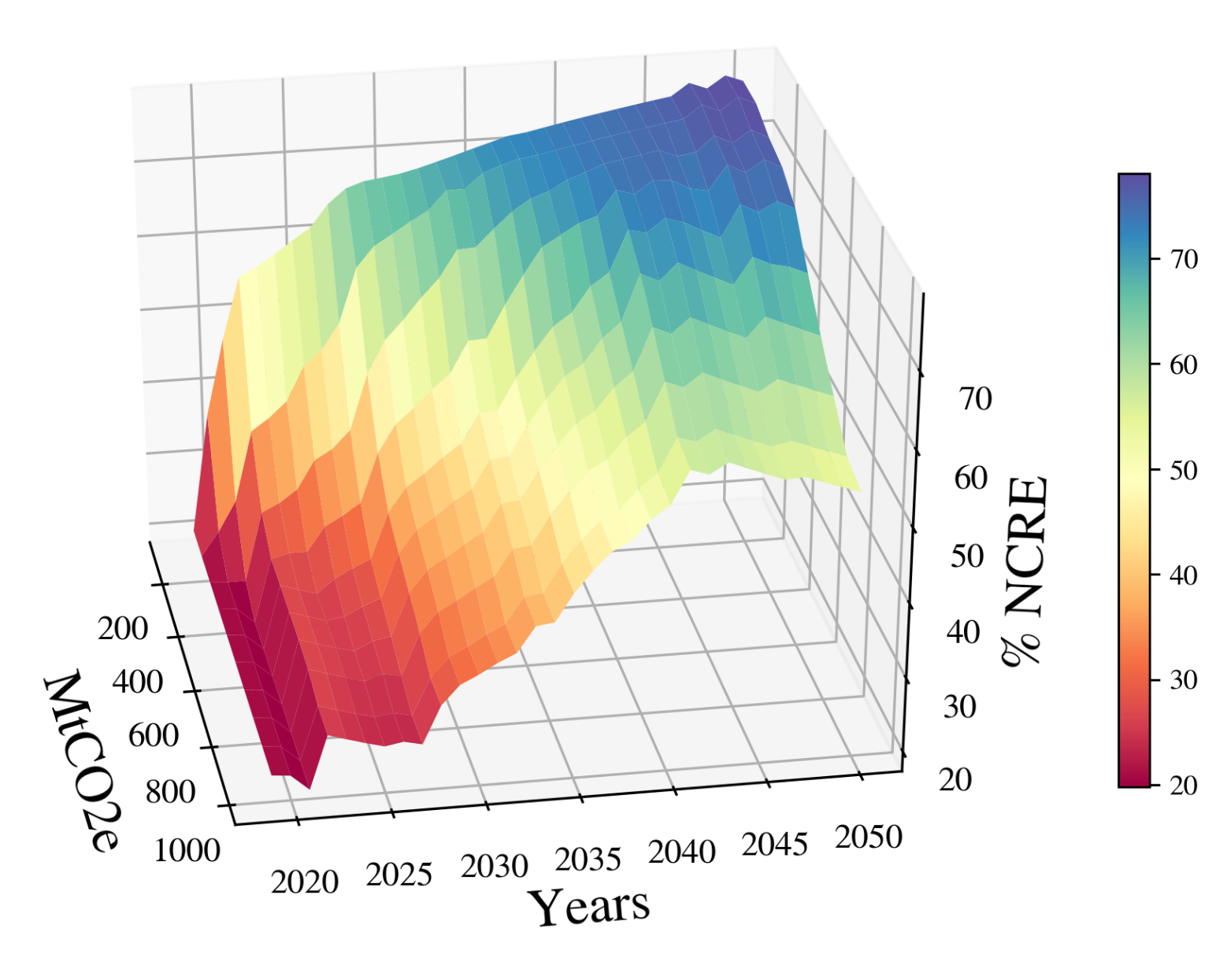}
 \caption{Percentage of power generation from NCRE sources in the period 2019-2050 at different carbon budgets. }
  \label{fig:ernc_percent_cap}
  \end{figure}
  
To quantify what carbon budgets and carbon permit prices are needed to meet the Chilean pledges, we examine the temporal evolution of the percentage of power production from NCRE sources, for each carbon budget level, as shown in Figure~\ref{fig:ernc_percent_cap}. We followed the usual definition of NCRE sources, i.e., biomass, solar, wind and mini hydro-RoR. Figure~\ref{fig:ernc_percent_cap} shows that, without further restrictions, a restrictive carbon budget could lead to more investment in NCRE technologies. For a carbon budget of 1000 MtCO$_2e$, by 2040 there is a 52\% of NCRE technologies in the generation mix. As explained in Section~\ref{sec:intro}, in the Chilean pledge, all hydro generation - hydro-dam, hydro-RoR and mini hydro-RoR were included as NCRE sources. This result leads to the conclusion that the Chilean pledge is an easy target, a conclusion also shared by other authors (e.g., see \cite{munoz2017aiming}). Our calculations show that 60\% of NCRE sources is achieved by 2041 with a carbon budget of around 700 MtCO$_2e$ (see Table \ref{tab:resultsTable}), and to hit the 70\% by 2050, the carbon budget should be close to 600 MtCO$_2e$.

\begin{table}[h]
    \centering
    \begin{tabular}{ l c c c c } 
 \hline
 Pledge & \multicolumn{4}{c }{Carbon budget (MtCO$_2e$)} \\
 & 100  & 500  & 700 & 1000\\ 
 \cline{2-4}
 \hline
Year in which coal is phased-out & 2022 & 2041 & -- & --  \\
Percentage ($\%$) of coal and gas in 2050 & 0 & 0 & 16 & 30  \\ 
Year  with 60\% NCREs & 2025 & 2038 & 2041 &--  \\ 
Year  with 70\% NCREs & 2034 & 2041 & -- &--  \\ 
Price of allowances (USD per tCO$_2e$) & 174 & 45 &35 &23 \\ 
 \hline
\end{tabular}
    \caption{Summary of results for the deterministic case}
    \label{tab:resultsTable}
\end{table}

Projections based on the analysis above show that the pledge regarding coal phase out is difficult to honor. Thus, it is necessary to investigate new mechanisms to accomplish the targets. 

The cap and trade configuration establish a permits market and no additional trade-restrictions for producers. Unrestricted access to markets is not a realistic setting since market access is ruled by revenue, capacity, caps or quotas among many other regulatory or institutional considerations. For instance, at a general level we analyzed caps that determine the size of the permits market. At an individual level, restricted access can take sophisticated (financial) shapes: options, collateral requirements, margin calls, personalized caps and many other financial instruments. In a competitive setting, (exogenous) restrictions as the one given by a quota were introduced by \cite{Cass1984,Cass2006}. Endogenous restrictions are analyzed in \cite{Cassetal2001,Carosietal2009} among others. Nevertheless, the case of endogenous investment restrictions was not included until \cite{cea-echenique_general_2018}.

\begin{figure}[ht!]
\centering
 \includegraphics[width=4.5in]{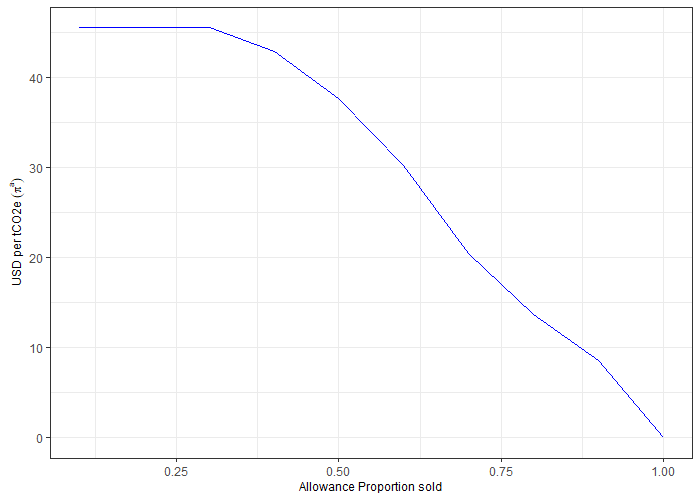}
 \caption{Allowance Price with respect to Allowance Proportion sold by Coal in $t=1$.}
  \label{fig:priceproportions}
  \end{figure}

We analyze the impact of an endogenous limit for the long position in the permits market. For a generator, the cap size (allowed emissions) is defined by a proportion of the permits bought in the first stage. The endogenous cap can be targeted to a specific technology (segmentation). The application of a targeted policy to segment the Coal-based technology implies in the elimination of the technology by 2035 if the proportion is 70\%. The effect of this targeted policy on prices induce a reduction in the price of allowances provided a proportion greater than 30\% (see Figure \ref{fig:priceproportions}). This is an interesting effect and beyond the scope of this work. Briefly, the price changes when all the allowances are bought by Coal. On the one hand, Coal needs more allowances to maintain operation when a greater proportion is required. On the other hand, once all the allowances are bought by Coal, other technologies seem to have benefits of a lower price of permits in the second stage.

\subsection{Impact of demand uncertainty on the Chilean pledges}\label{sec:scenarios}
\begin{figure}[ht]
    \begin{subfigure}[t]{0.5\linewidth}
    \includegraphics[width=\textwidth]{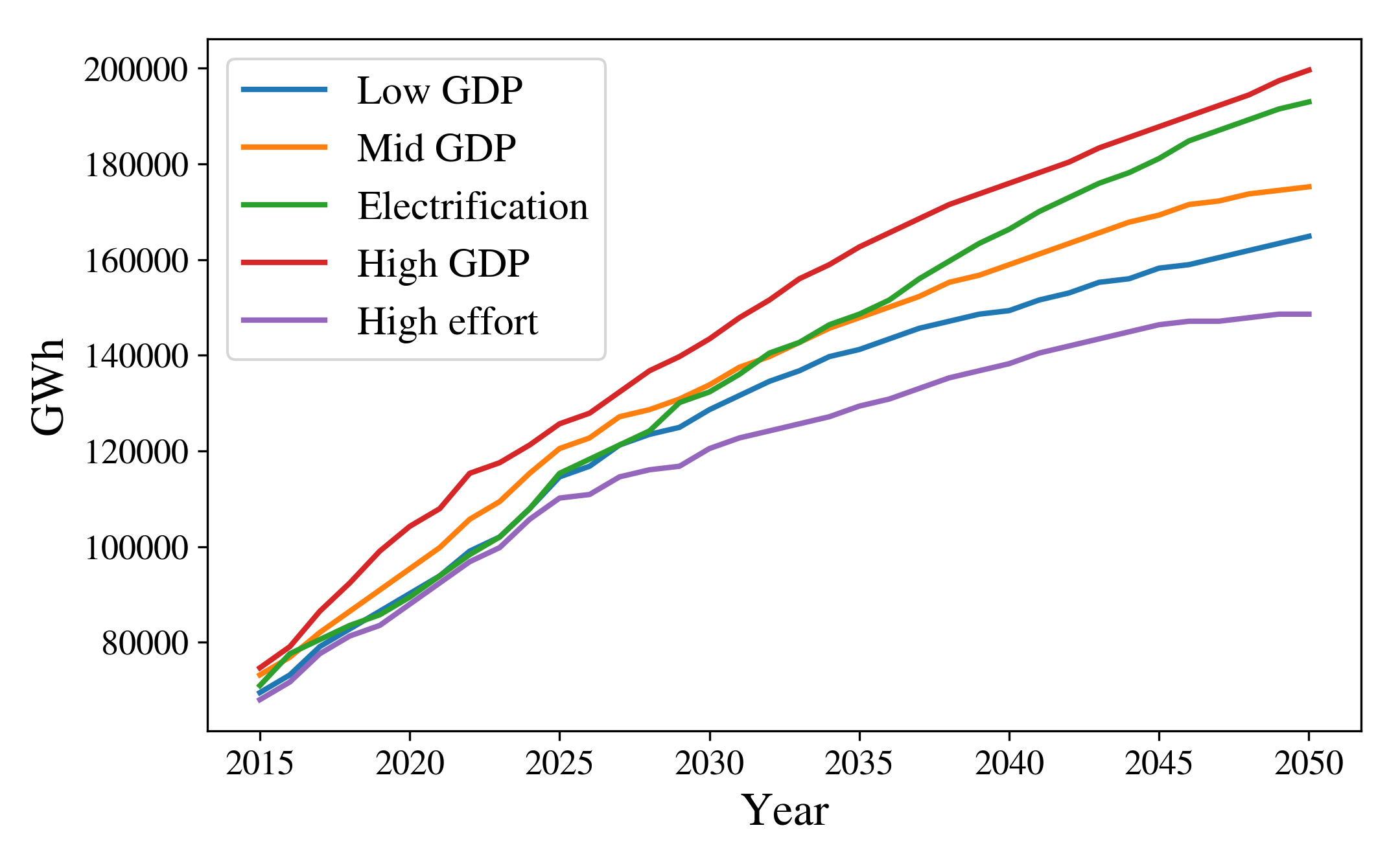}
    \caption{}
  \label{fig:demand_scn}
  \end{subfigure}
  \begin{subfigure}[t]{0.5\linewidth}
    \includegraphics[width=\textwidth]{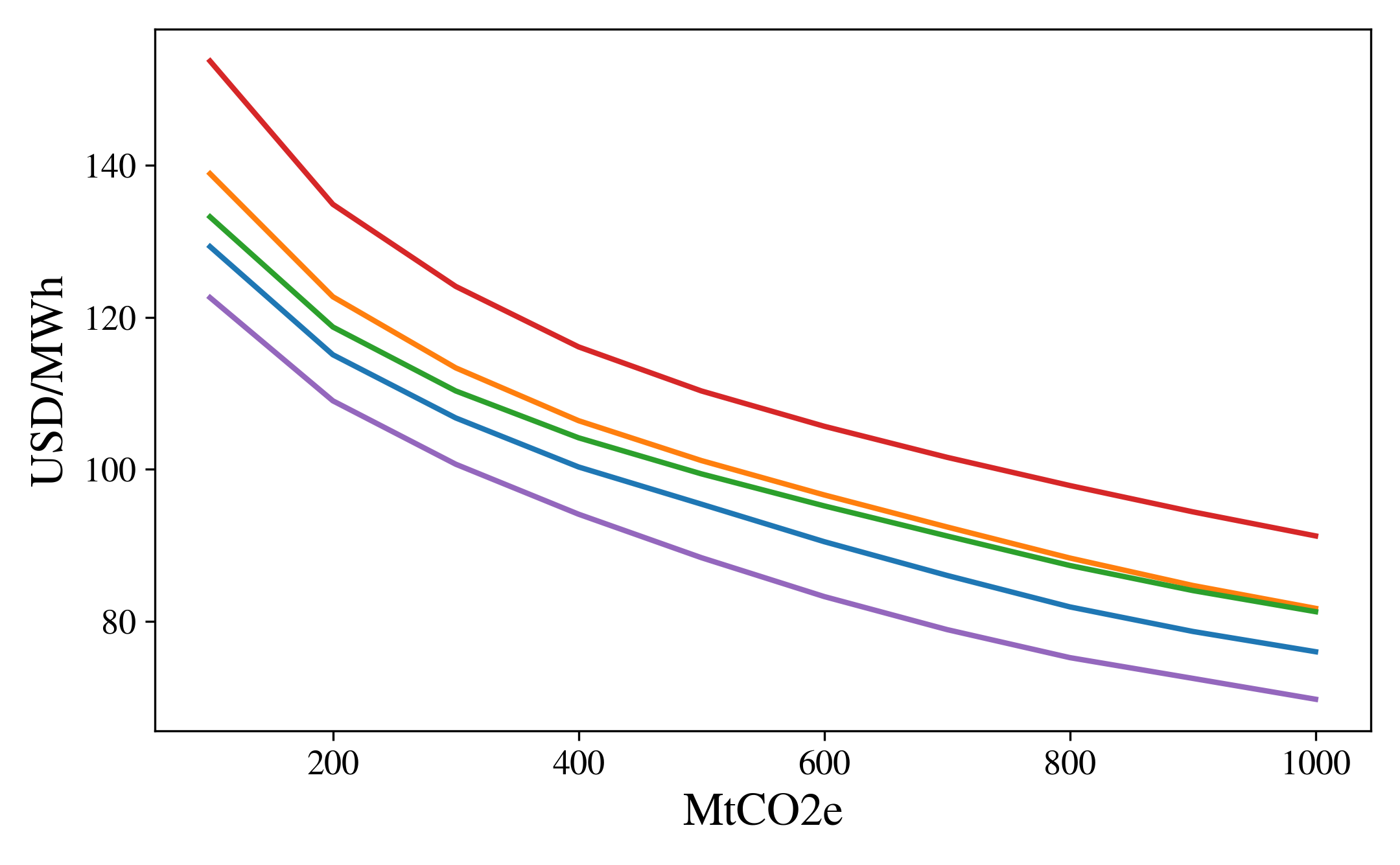}
    \caption{ }
  \label{fig:mean_price}
  \end{subfigure}
  \begin{subfigure}[t]{0.5\linewidth}
    \includegraphics[width=\textwidth]{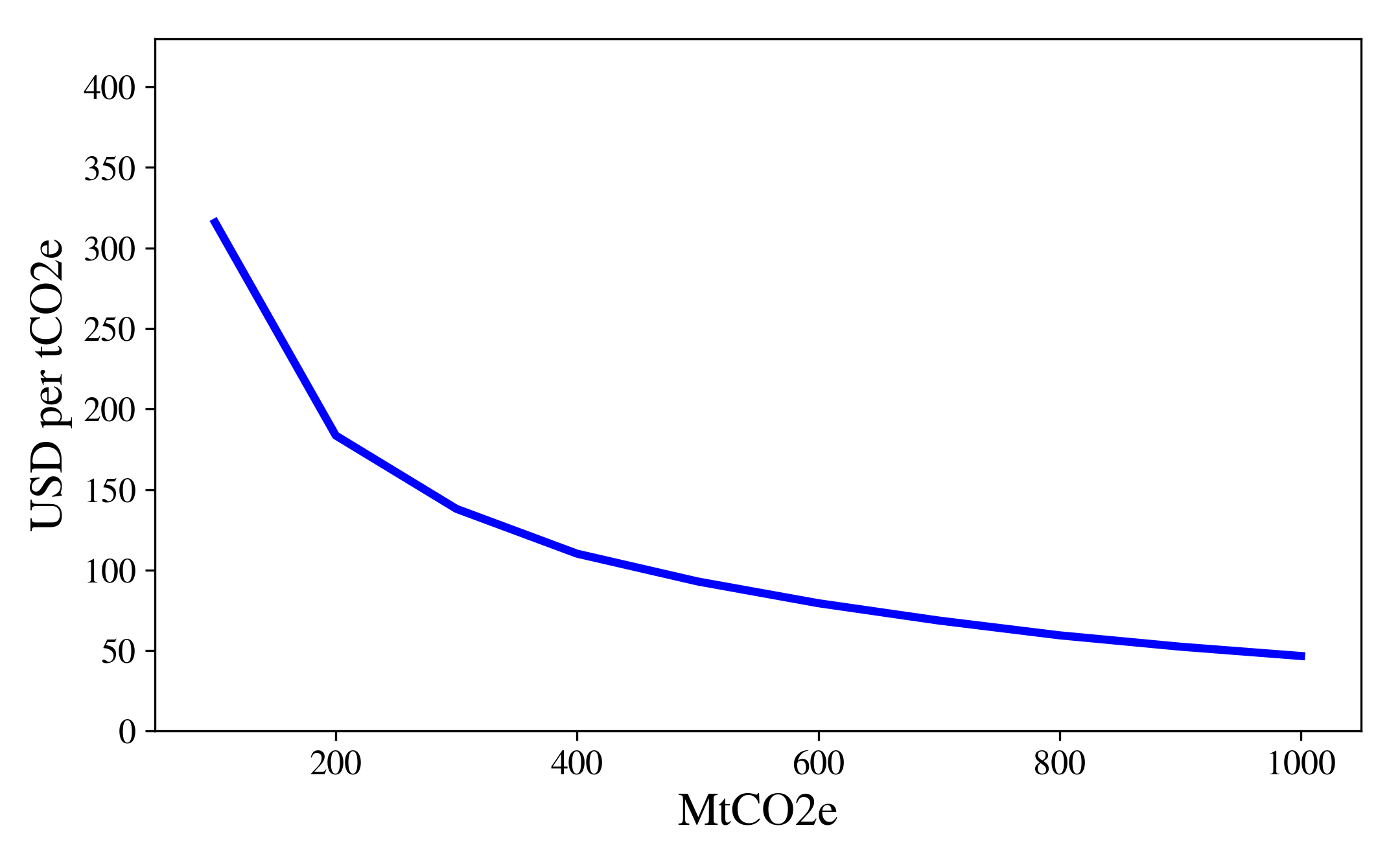}
    \caption{  }
  \label{fig:price_allo}
  \end{subfigure}
    \begin{subfigure}[t]{0.5\linewidth}
    \includegraphics[width=\textwidth]{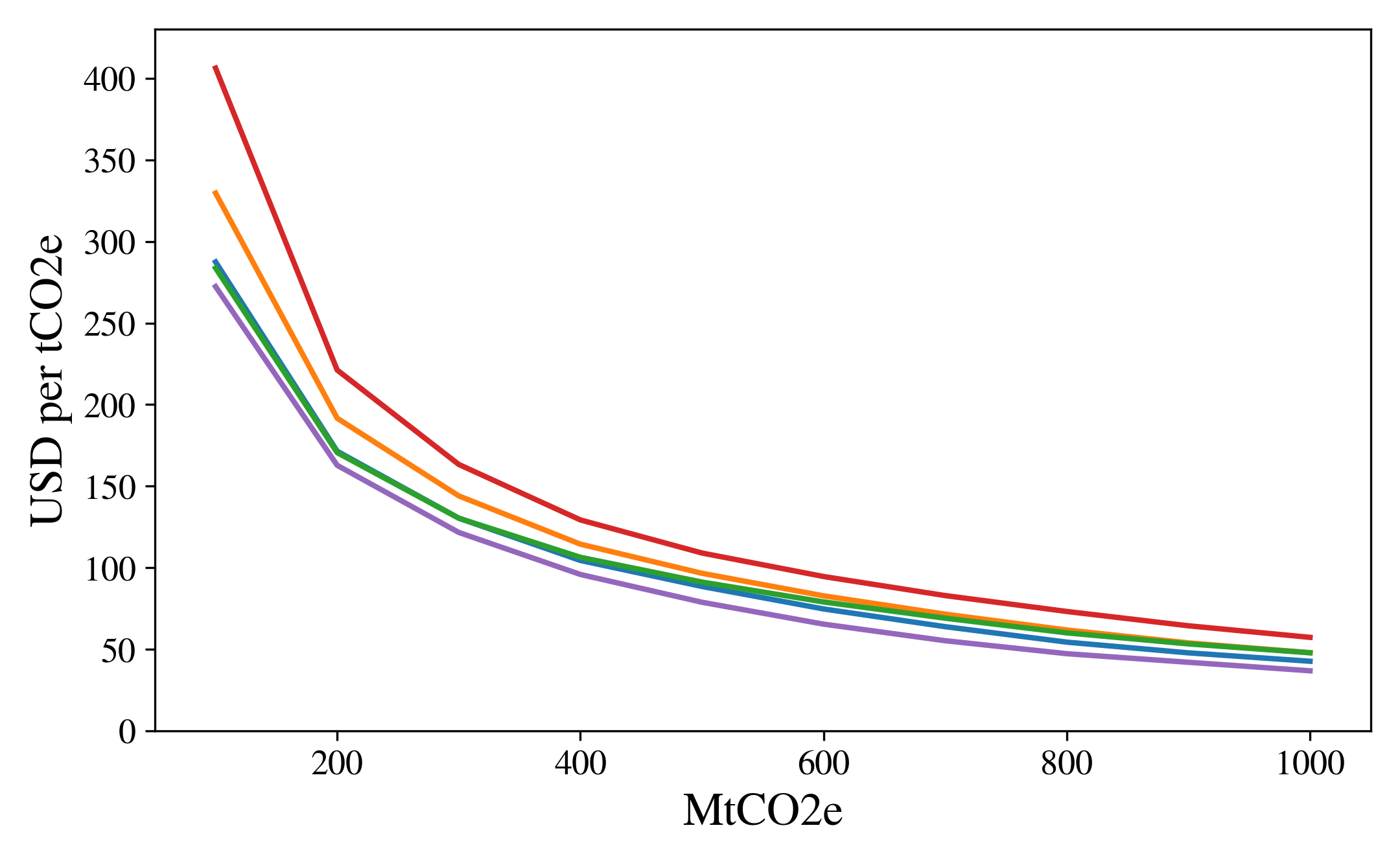}
    \caption{}
  \label{fig:price_trade}
  \end{subfigure}\quad
   \caption{Panel (a): Demand scenarios. Panel (b): Electricity prices per MWh, for each carbon budget value and different demand scenarios. Panel (c): Price of allowances per ton CO$_2e$ for each carbon budget. Panel (d): Allowance trading price per ton CO$_2e$ for each carbon budget and different demand scenarios. Color legend for Panels (b) and (d) are the same as in Panel (a).}
\end{figure}

We now consider the case where there is demand uncertainty in future periods. The stochastic demand creates uncertainty with respect to how many allowances generators should get during the first stage, hence, creating different carbon price patterns. We consider five demand scenarios, based on projections from Chile's roadmap towards 2050
\cite{HojadeRuta}\footnote{Note that this is a different entity than the CNE, which provided the real historical demand and its projection used in the case of the deterministic analysis presented in Section \ref{sec:deterministic}}. We also assume that each scenario is equally likely, $Pr(\omega)=1/5$ for each $\omega\in\Omega$. The demand scenarios considered here are \textit{Low, Mid and High GDP, Electrification and High Effort}, as shown in Figure~\ref{fig:demand_scn}. The Electrification scenario assumes a replacement of fuels for electricity in tertiary sectors, such as heating, industries and transportation. The high effort scenario assumes energy efficiency policies in action, resulting in a significant reduction on electricity consumption as compared to the other scenarios. 

\small
\begin{table}[h]
    \centering
    \begin{tabular}{ l| c c c   } 
\hline
Scenario	    &	BAU    	&	1000 (\% increase from BAU)	    &	100 (\% increase from BAU)	        \\
\hline
Low GDP	        &	29.47	&	51.52 (0.74)	&	57.20 (0.94)	\\
Mid GDP 	    &	33.73	&	59.80 (0.77)	&	62.34 (0.84) 	\\
Electrification	&	41.96	&	66.79 (0.59)	&	70.65 (0.68)	\\
High GDP	    &	43.13	&	72.72 (0.68)	&	73.96 (0.71)	\\
High Effort	    &	22.93   &	41.08 (0.79)	&	49.79 (1.17)	\\
\hline
\end{tabular}
    \caption{New capacity investment for distinct demand scenarios and emissions budget.}
    \label{tab:InvestmentStoch}
\end{table}
\normalsize

\smallskip
Figure~\ref{fig:mean_price} shows the resulting mean price of electricity over the period 2019-2050 for different carbon budgets and for each of the different demand scenarios. As expected, higher demand scenarios result in higher mean electricity prices (31 \% higher for a carbon budget on 1000 MtCO$_2e$), which are further increased with more restrictive carbon budgets. In fact, the increment of the mean electricity price when the carbon budget is reduced from 1000 MtCO$_2e$ to 100 MtCO$_2e$ is equal to 68\% ($\$91/$MWh to $\$154/$MWh) and 76\% (\$70$/$MWh to $\$123/$MWh) for the High GDP and High effort scenarios, respectively. Interestingly, the Electrification and Mid GDP scenarios have a similar mean electricity price for less restrictive carbon budgets (600-1000 MtCO$_2e$). Such behavior is explain by the offset effect observed in the demand patterns between these two scenarios before and after 2030.
\smallskip

We also examine the observed allowance price $\pi^a$, as shown in Figure \ref{fig:price_allo}. Note that the allowance price is not indexed by the scenarios $\omega$, however, the resulting price depends on the realization of each scenario $\omega$ (Equation~\ref{eq:pi_a_appendix}). In fact, facing uncertainty, emitting units must now acquire emissions permits that would allow them to generate electricity for each possible realization of demand.  Figure~\ref{fig:price_allo} shows such price for different carbon budget scenarios. We observed prices ranging between 46 USD per tCO$_2e$ for a budget of 1000 MtCO$_2e$ to over 300 USD per tCO$_2e$ when a carbon budget of 100 MtCO$_2e$ is considered. Note that the prices obtained here are in good agreement with what authors have claimed is a suitably, efficient price for carbon emissions (for instance, see Table~1 from \cite{feijoo2019climate}). High prices are mainly due to the impact of the high demand scenarios considered here. In fact, Figure \ref{fig:price_trade} shows the carbon permit trading price. It is clear from the figure that the value that generators see on an additional carbon permit is much higher in the case of High GDP (high demand), with carbon trading prices surpassing \$400 per tCO$_2e$.  
\smallskip

\begin{figure}[ht]
\centering
    \includegraphics[width=5.5in]{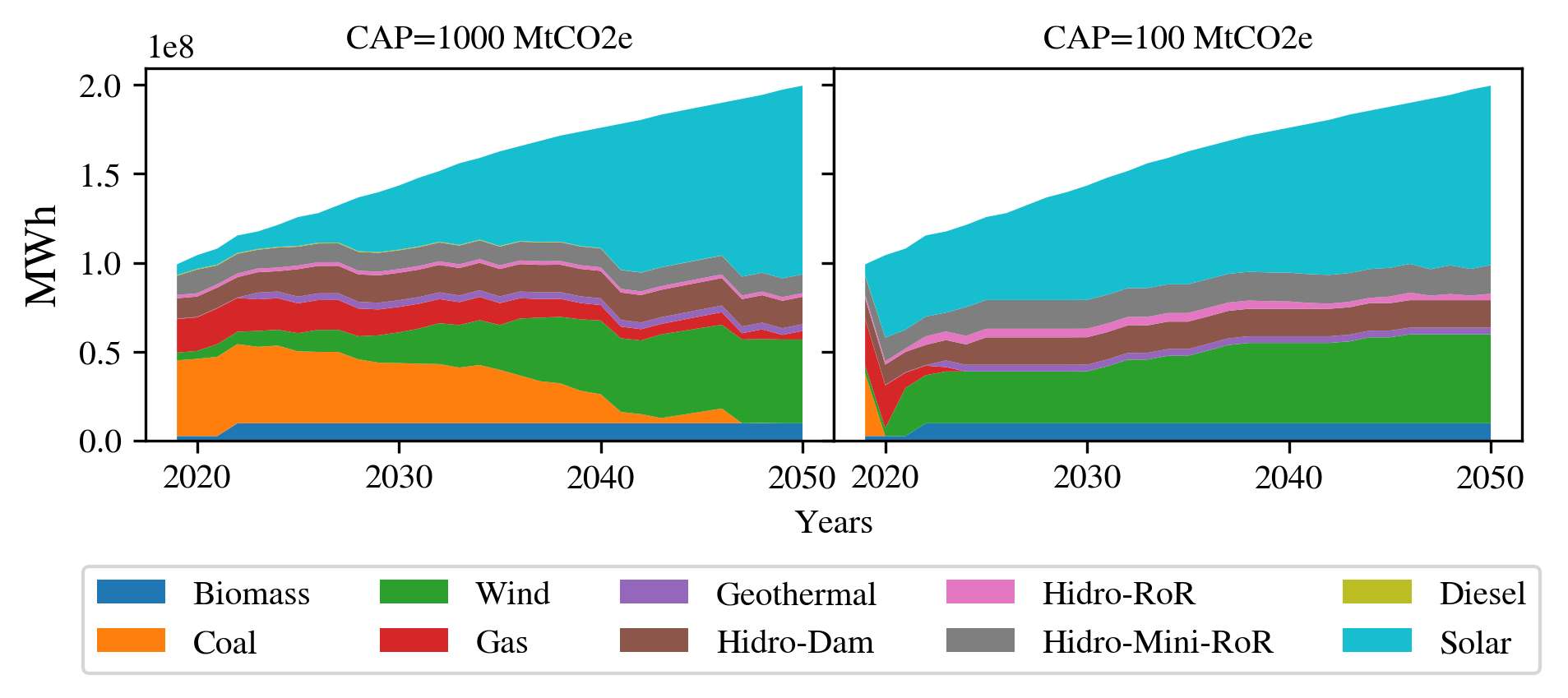}
\caption{Production levels per technology for the period 2019-2050 for different carbon budgets.  Demand scenario : High GDP.}
  \label{fig:comp_100_1000}
\end{figure}

Table \ref{tab:InvestmentStoch} shows the cumulative capacity investments over the period 2019-2050 for a BAU case (no carbon budget with uncertainty considered) and for the two extremes of the carbon budget, 100 and 1000 MtCO$_2e$. For the BAU case, we observe that the two scenarios with higher demand also have the higher amount of new capacity additions. The High GDP scenario installed 43 GW of capacity (highest), 88\% more than the case of the High effort scenario (lowest demand). In fact, we do obtain low levels of additions in coal power plants (less than 1 GW) in all scenarios but in the High effort and Low GDP, the two cases with lowest demand. However, when a carbon budget is considered (even for high levels of a carbon budget), we no longer observe additions in coal technologies. 
\smallskip
Interestingly, the scenario of High GDP has similar levels of new investment across different levels of carbon budget constraints. For a restrictive carbon budget (100 MtCO$_2e$), we obtain almost 74 GW of new capacity for the High GDP scenario (71\% increase from the BAU case) compared to 72.72 GW in the case of 1000 MtCO$_2e$, only a 3\% difference despite the large differences in the emissions that are allowed. This suggests that under low levels of carbon constraint and a high future demand, renewable energy will likely prevail due to cost increase faced by fossil fuel technologies from the cost of accessing allowances. Also, in a high demand scenario, a less restrictive carbon budget is tight due to the lower historical demand level used to estimate such budget. This is clear from Figure \ref{fig:comp_100_1000}, where we see that for the same demand level, a low carbon budget faces out coal and gas (main fossil fuel base emitters) right after 2020, whereas a larger budget faces out coal towards 2050, with a low share of natural gas still remaining. However, since the demand level in both cases is the same, we obtain similar levels of renewable investment needed to satisfy demand by 2050. The slightly lower investment in the 1000 MtCO$_2e$ scenario results from the small share of natural gas-fired power plants that are still in place. 
\smallskip

\begin{figure}[ht!]
\centering
 \includegraphics[width=4.5in]{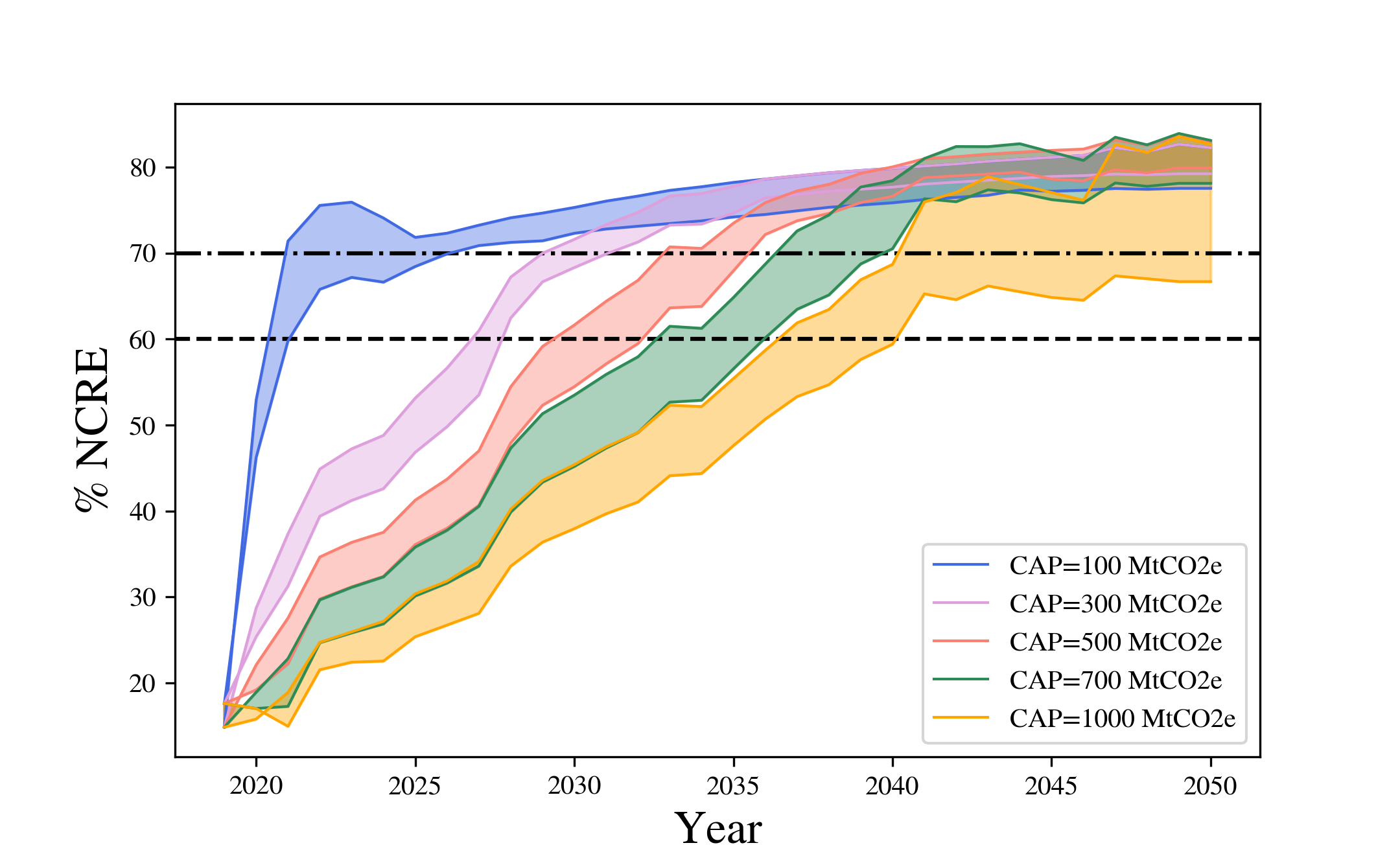}
 \caption{Percentage of power generation from NCRE sources for different carbon budgets. The shaded areas represent the range between the highest demand scenario (High GDP) and the lowest (High Effort). The dash and dash-dot lines mark the level at 60\% and 70\% of NCRE sources.}
  \label{fig:ernc_percent_shadows}
  \end{figure}

Finally, Figure \ref{fig:ernc_percent_shadows} shows, for a set of carbon budgets, the range of NCRE share over time based on the demand scenarios (range defined between the High GDP and High effort scenarios, as high and low demand). Clearly, a stringent carbon budget easily reaches all the pledges before 2030, independently of the demand scenario that is realized. This is also true for other budgets, however, the timing at which the pledge is met changes significantly. In the case of a lax carbon budget cap, the 70\% pledge is met after 2040 for some cases of the realization of demand. However, when the 60\% cap is analyzed, we observed that the pledge is always met. As in the case of the investment additions describe above in Table \ref{tab:InvestmentStoch}, the demand scenarios here are higher than the demand studied in the deterministic case. Hence, under cases of higher demand scenarios, the technology change faced by NCRE (mainly solar and wind) generates an intuitive incentive for increased shares of such renewable energy sources. In fact, a lax carbon budget, for the highest demand scenario only (High GDP), faces out coal by 2049-2050. All other scenarios of demand still have a significant share of coal (and natural gas) generation. 

\section{Conclusions}\label{sec:concl}

This paper explores the capacity investment needs of the Chilean electric sector required to comply with its national and international energy and climate commitments. We impose a cap-and-trade system that allows to price carbon and to measure what is the actual remaining carbon budget. Each electric sector technology is independently modeled as a two stage stochastic capacity expansion and production problem. The market equilibrium is obtained by solving a mixed complementarity problem that correspond to the optimality condition of each electric technology and the cap-and-trade auctioneer. Results show that, under a deterministic demand base projection, if the commitment of phasing out coal-based power plants is met, then such solution leads to meet NCRE targets in the near and long term. However, the opposite is not true. Meeting the required percentage of production of electricity from green energy sources does not necessarily mean that coal plants are removed from the electricity mix. A similar result is observed when considering an uncertain future demand. 

Particularly, we find that in order to phase out coal by 2040 and to achieve 70\% of NCRE by 2050, the remaining carbon budget (as used in the cap-and-trade policy) is close to 500 MtCO$_2e$, almost half of the current Chilean estimates (slightly lower than 1000 MtCO$_2e$, see Section \ref{sec:cap}). Hence, it is necessary to focus in carbon budget targets that are more stringent than the current one pledge by Chile. Also, our estimates suggest that carbon prices that are aligned with the energy targets are close to 50 USD per tCO$_2e$ or above, far from the current tax policy implemented in Chile. 



Market segmentation may help in the achievement of promises. Precisely, by defining endogenous caps on trading permits when the allowances are already distributed, coal phase out could be attained in 2035. Nevertheless, there is still the need of a more stringent carbon budget. There are relations of the endogenous caps with other variables that need further analysis, e.g., as the impact on prices, the role of inter-temporal substitution or the cost of stranded assets. In this context, it is important to account for possible financial burden. Generally, the cost of stranded power plants are significant for developing economies, such as Chile or Latin America. Such a cost has been estimated to be between 37-90 billion USD over the next 30 years in the Southern American region \cite{feijoo2019Stranded}. Hence, it becomes extremely important to address the implications of policies and to improve the mitigation strategies. Some authors (e.g., \cite{Johnson2015} and \cite{Clark2014}, and references therein) have shown that the inclusion of Carbon Capture and Storage (CCS) technologies can be an effective strategy to reduce the economic impact of stranded assets, particularly, when more serious carbon policies are enacted.


\newpage
\begin{appendices}
\appendix
\section{MCP Formulation}\label{ap:mcp}
\setcounter{equation}{0}
\numberwithin{equation}{section}
The Producer problem is 


\small
\begin{align}
\min_{(x_i,Q_i,A_i,P_i,V_i)\in \mathbb{X}}  & f_i \big( \pi^d(0),Q_i(0)\big)+ A_i \pi^{a} + I_i x_i(0) \nonumber \\ 
& + \sum_{\omega} Pr(\omega)   \Bigg[ \sum_{t>0} \frac{1}{(1+R)^t} \Big[ TC_i(t)\cdot f_i \big( \pi^d(t,\omega),Q_i(t,\omega) \big) \nonumber\\
&  + TCR_i(t) \cdot I_i\cdot x_i(t,\omega) \Big] + \pi^v(\omega)\cdot \big(P_i(\omega)-V_i(\omega)\big) \Bigg]   \label{eq:prodA}
\end{align}
\begin{align}
&\textrm{subject to \ } \nonumber\\
    &\Big(CF_i \cdot\tau\Big)  \Bigg[\bar{Q}_i + x_i(0)+\sum_{t^{\prime}<t-lag_i} x_i(t^\prime,\omega) \Bigg] - Q_i(t,\omega) & \geq 0  & \qquad \forall \quad i,\omega, t  > 0   & \quad (\alpha_{i,\omega,t})      \label{eq:c1A}\\
    &\Big(CF_i\cdot\tau \Big)\bar{Q_i}-Q_{i}(0)                                                                                 & \geq 0  & \qquad \forall \quad i                  &  \quad (\kappa_i) \label{eq:c2A} \\
    &RP_i - \bar{Q}_i - x_i(0) - \sum_{t > 0} x_i(t,\omega)                                                                     & \geq 0  & \qquad \forall \quad i,\omega           &   \quad (\psi_{i,\omega})\label{eq:c3A} \\
    &A_{i} -V_i(\omega)                                                                                                         & \geq  0 & \qquad \forall \quad i,\omega           & \quad (\beta_{i,\omega}) \label{eq:c4A}\\
    &A_{i} + (P_i(\omega) - V_i(\omega))-\sum_{t>0}Q_i(t, \omega)\cdot \varepsilon_{i}-Q_i(0)\varepsilon_{i}           & \geq  0 & \qquad \forall \quad i, \omega          & \quad (\gamma_{i,\omega}) \label{eq:c5A} \\
    &Q_i(0) & \geq  0 & \quad & \quad (\lambda_i)\\
    &Q_i(t, \omega) & \geq  0   & \qquad \forall  \quad \omega, t >0 & \quad (\delta_{i,\omega,t})\\
    &x_i(0) & \geq  0 & \quad & \quad (\xi_i)\\
    &x_i(t, \omega) & \geq  0   & \qquad \forall  \quad \omega, t >0 & \quad (\varphi_{i,\omega,t})
  \end{align}
\normalsize
  
Using this problem definition and nomenclature defined in Section~\ref{sec:model}, we can write the Lagrangian function for the producer's problem of each $i \in \{ 1,...,N\}$ as follows:
\small{
\begin{align}
&\mathcal{L}_i(x_i,Q_i,A_i,P_i,V_i) = f_i \big( \pi^d(0),Q_i(0)\big)+ A_i \pi^{a} + I_i x_i(0)  +& \nonumber \\ 
&\sum_{\omega} Pr(\omega)\Bigg[ \sum_{t>0} \frac{1}{(1+R)^t} \Big[ TC_i(t)\cdot f_i \big( \pi^d(t,\omega),Q_i(t,\omega) \big) + TCR_i(t) \cdot I_i\cdot x_i(t,\omega) \Big] + \pi^v(\omega)\cdot \big(P_i(\omega)-V_i(\omega)\big) \Bigg]   + &\nonumber \\
&\kappa_{i}\Big[Q_i(0) -  \big(CF_i\cdot\tau \big)\bar{Q}_i \Big] +\sum_{\omega,t>0} \alpha_{i,\omega,t}\Bigg[Q_i(t,\omega) - \big(CF_i \cdot\tau\big) \big(\bar{Q}_i + \sum_{t^{\prime} \leq t } x_i(t,\omega) + x_i(0) \big)\Bigg] + & \nonumber \\ &\sum_{\omega}\beta_{i,\omega}\Big[V_i(\omega)-A_i \Big] + \sum_{\omega}\gamma_{i,\omega} \Big[-A_{i} - P_{i}(\omega) + V_i(\omega) +\sum_{t>0} Q_i(t,\omega) \varepsilon_{i} + Q_i(0)\varepsilon_{i}\Big] - \sum_{\omega, t>0}\delta_{i,\omega,t} Q_i(t,\omega) + & \nonumber \\ &\sum_{\omega}\psi_{i,\omega} \Big[  \bar{Q}_i+ x_i(0) + \sum_{t > 0} x_i(t,\omega) - RP_i \Big] - \lambda_{i}\Big[Q_{i}(0)\Big] - \sum_{\omega, t>0}\varphi_{i,\omega,t} x_i(t,\omega) - \xi_i x_i(0)& \label{eq:lagrange}
\end{align}
}

where $\alpha_{i,\omega,\tau}$, $\kappa_i$, $\beta_{i,\omega}$, $\gamma_{i,\omega}$, $\delta_{i,\omega,t}$  $\lambda_i$, $\varphi_{i,\omega,t}$, $\xi_i$ and $\psi_{i,\omega}$ are the lagrange multipliers of the constraints.

\smallskip
Based on this lagrangian function, we use the Karush-Kuhn-Tucker optimality conditions to formulate the problem as an Mixed Complementarity Problem. We will use the compact notation using the $\perp$ operator denoting the inner product of two vectors equal to zero.

\smallskip
The KKT conditions of the producer's problem are:

\footnotesize{
\begin{align}
    & 0 \leq I_i  + \sum_{\omega}\psi_{i,\omega} -\sum_{\omega, t>0} \alpha_{i,\omega,t} \perp  x_i(0) \geq 0  \qquad \forall \  i \\
    & 0 \leq Pr(\omega) \Bigg[\frac{1}{(1+R)^t}TCR_i(t) \cdot I_i \Bigg] - \sum_{t> t\prime}\alpha_{i,\omega,t} ( CF_i \cdot \tau)+ \psi_{i,\omega} \perp x_i(t,\omega) \geq 0  \qquad  \forall \  i, \omega, t> 0\\
    &  0 \leq  \big(a_{i}+b_i Q_{i}(0)\big)-\pi^d(0) + \kappa_i  + \sum_{\omega} \gamma_{i,\omega}\varepsilon_i \perp  Q_{i}(0) \geq 0 \qquad \forall \  i  \\
    &  0 \leq  Pr(\omega)  \frac{1}{(1+R)^t} \bigg( TC_i(t) \big(a_{i}+b_i Q_i(t,\omega)\big ) -\pi^d(t,\omega) \bigg) + \alpha_{i,\omega,\tau} + \gamma_{i,\omega} \varepsilon_{i} \perp Q_i(t,\omega) \geq 0 \qquad  \forall \ i, \omega, t > 0\\
    & 0 \leq \pi^{a} - \sum_{\omega}\beta_{i,\omega} - \sum_{\omega}\gamma_{i,\omega} \perp A_i \geq 0 \qquad \forall \  i  \label{eq:pi_a_appendix}\\
    & 0 \leq -Pr(\omega) \pi^v(\omega) + \beta_{i,\omega}  + \gamma_{i,\omega} \perp V_i(\omega) \geq 0 \qquad \forall \  i, \omega \\
    &  0 \leq Pr(\omega) \pi^v(\omega) -\gamma_{i,\omega} \perp P_{-i}(\omega) \geq 0 \qquad \forall  \ -i,\omega \\
    & 0 \leq \big(CF_i \cdot \tau \big) \Bigg[\bar{Q}_i + \sum_{t\leq t^{\prime}} x_i(t,\omega) + x_i(0) \Bigg] - Q_i(t,\omega)  \perp \alpha_{i,\omega,\tau} \geq 0 \qquad \forall \ i, \omega, t  > 0\\
    & 0 \leq \Big(CF_i\cdot\tau \Big)\bar{Q}_i(0)-Q_{i}(0) \perp \kappa_i \geq 0 \qquad \forall \ i \\
    & 0 \leq  A_{i} - V_i(\omega) \perp \beta_{i,\omega} \geq 0 \qquad \forall  \ \omega \\
    & 0 \leq  A_{i} + P_{i} (\omega) - V_i(\omega) - \sum_{t>0} Q_i(t,\omega) \varepsilon_{i} -Q_i(0)\varepsilon_{i} \perp \gamma_{i,\omega} \geq 0 \qquad \forall \ i, \omega\\
    & 0 \leq  RP_i - \bar{Q}_i - x_i(0) - \sum_{t>0} x_i(t,\omega) \perp \psi_{i,\omega} \geq 0 \qquad \forall \ i,\omega 
\end{align}
}

\normalsize
For the auctioneer, the Lagrangian function is 

\begin{equation}
    \mathcal{L}(\theta)= -\theta \pi^{a} + \mathcal{F}(\theta) - \eta (\phi^{-1}(M) \sigma+ \mu - \theta)
\end{equation}

where $\eta$ and $\zeta$ are the Lagrange multipliers. The KKT conditions for the auctioneer problem are

\begin{align}
    & 0 \leq -\pi^{a}+ \frac{\partial \mathcal{F}(\theta)}{\partial \theta} +\eta \perp \theta \geq 0 \\
    & 0 \leq \phi^{-1}(M) \sigma+ \mu - \theta  \perp \eta \geq 0
\end{align}

\end{appendices}

\newpage
\bibliographystyle{plainnat}
\bibliography{Main}

\end{document}

%% file: resources.tex
\makeatletter
\newcommand*{\inlineequation}[2][]{%
  \begingroup
    \refstepcounter{equation}%
    \ifx\\#1\\%
    \else
      \label{#1}%
    \fi
    \relpenalty=10000 %
    \binoppenalty=10000 %
    \ensuremath{%
      #2%
    }%
    ~\@eqnnum
  \endgroup
}
\makeatother

\usepackage{amssymb,amsthm,color,enumerate,graphicx,mathrsfs,xspace,algorithm,algorithmicx,algpseudocode,booktabs,adjustbox,soul,threeparttable,pdflscape,afterpage,longtable,comment,multirow,etoolbox,subcaption}
\usepackage[usenames,dvipsnames]{xcolor}
\usepackage[sort&compress,comma, numbers]{natbib}
\usepackage[colorlinks=true, allcolors=blue]{hyperref}
\usepackage[nameinlink]{cleveref}
\usepackage[inline]{enumitem}

\renewcommand{\bar}{\overline}

\colorlet{rosso}{red!90!white}
\definecolor{green}{RGB}{45, 204, 53}
\definecolor{blue}{RGB}{59, 42, 247}

\allowdisplaybreaks
\crefname{lemma}{Lemma}{Lemmata}
\crefname{theorem}{Theorem}{Theorems}
\crefname{claim}{Claim}{Claims}
\crefname{algorithm}{Algorithm}{Algorithms}
\crefname{equation}{}{}
\crefname{Def}{Definition}{Definition}
\crefname{Cla}{Claim}{Claim}
\crefname{Corr}{Corollary}{Corollaries}
\crefname{remark}{Remark}{Remarks}
\crefname{Ex}{Example}{Examples}
\crefname{figure}{Figure}{Figures}
\crefname{section}{Section}{Sections}
\crefname{table}{Table}{Tables}
\crefname{enumi}{Statement}{Statements}
\crefname{line}{Step}{Steps}

\definecolor{mycolor}{RGB}{200,0,0}

\newtoggle{ExtendedVersion}  